\documentclass[12pt]{article}
\pdfoutput=1
\usepackage{jheppub}
\usepackage{amsfonts}
\usepackage{amsmath,amssymb} 
\usepackage{graphicx} 
\usepackage[dvipsnames,x11names]{xcolor} 

\usepackage[nottoc,notlof,notlot]{tocbibind} 
\usepackage[titles]{tocloft} 
\usepackage{hyperref}
\usepackage{cleveref}
\usepackage{float}
\usepackage{subcaption}
\numberwithin{equation}{section} 

\title{\textbf{Phase transition and chaos in charged SYK model}}

\author{Nilakash Sorokhaibam}\emailAdd{nilakash@niser.ac.in}

\affiliation{
National Institute of Science Education and Research, HBNI, Bhubaneswar 752050, Odisha, India}

  \abstract{We study chaotic-integrable transition and the nature of quantum chaos in  SYK model with chemical potential. We use a novel numerical technique to calculate the partition function explicitly. We show the phase transition in the presence of large chemical potential. We also show that a mass-like term consisting of two fermion random interaction ($q=2$ SYK term) does not give rise to a sharp transition. We find that turning on the chemical potential suppresses the Lyapunov exponent in the chaotic phase exponentially.}

\begin{document}
\maketitle

\section{Introduction and Summary}
\label{sec:intro}
The Sachdev-Ye-Kitaev (SYK) model is a quantum system of many fermions ($N$ in number, $N$ is large) with random all-to-all interaction \cite{Sachdev:1993,Kitaev:2015}. It  has been a subject of great interest in the last few years \cite{Polchinski:2016xgd,Maldacena:2016hyu,Bulycheva:2017ilt,Kourkoulou:2017zaj,Dhar:2018pii,Numasawa:2019gnl}. The model has many remarkable properties. It does not have any quasi-particle excitations. The gaps in the spectrum are exponentially suppressed in $N$. It flows to a conformal theory in deep infrared. It also saturates the quantum chaos bound of \cite{Maldacena:2015waa}. All these properties point to the existence of a bulk dual of the theory. There has been many proposals and other related works on the gravity side \cite{Jevicki:2016bwu,Mandal:2017thl,Das:2017pif,Kitaev:2017awl,Das:2017wae,Gaikwad:2018dfc,Nayak:2018qej,Moitra:2019bub,Moitra:2019xoj}.

The original SYK model is a model with Majorana fermions. If one considers SYK model with complex fermions, one can turn on the mass term in the Hamiltonian or consider a thermal state with chemical potential turned on. With chemical potential turned on, a first order phase transition has been observed \cite{Azeyanagi:2017drg}. The high temperature phase is chaotic while the low temperature phase is integrable (non-chaotic). Henceforth, the two phases will be called chaotic phase and integrable phase. The integrable phase is effectively described by a weakly interacting massive theory. In this phase, the Lyapunov exponent is also practically zero.
This phase transition is like Hawking-Page transition between black hole phase and thermal AdS phase \cite{Hawking:1982dh}.

In this paper we study the phase transition in more details.
We calculate the partition function explicitly by evaluating the determinant in the expression of the partition function. We also study the chaos dynamics in the presence of mass term in the Hamiltonian and in a state with the chemical potential turned on.

We will mainly work with 4-fermion $(q=4)$ interaction. Consider the two Hamiltonians
\begin{equation}
H_{SYK}=\sum_{i,j,k,l=1}^{N}j_{4,ij;kl}\Psi_i^{\dagger}\Psi_j^{\dagger}\Psi_k\Psi_l, \qquad \tilde{H}_{SYK}=H_{SYK}+\mu\sum_{i=1}^{N} \Psi_i^{\dagger}\Psi_i\
\label{twoH}
\end{equation}
The couplings $j_{4,ij;kl}$ are random numbers drawn from a Gaussian distribution. For both the Hamiltonians, $Q=\sum_{i} \Psi_i^{\dagger}\Psi_i$ is a conserved charge. So,we can consider thermal states with non-zero chemical potential of this charge. The relation between charge $Q$ (the expectation value) and the chemical potential is given in \cite{Sachdev:2015efa} for a fixed $\mu\beta$ at low temperature limit $\beta \to \infty$ and $\mu\to 0$. Mass and chemical potential are same if one is calculating partition function using imaginary time path integration. The imaginary time Schwinger-Dyson (SD) equations are solved numerically using an iterative method. The solution is then used to calculate the partition function. To prepare low temperature states, one has to gradually cool down the system numerically. One can also heat up the system after the cooling process.
The details are given in section \ref{phasetrans}.

We have also looked for phase transition in another related system where instead of the mass term there is a ($q=2$) SYK interaction term. The Hamiltonian is
\begin{equation}
H_{SYK}+\sum_{i,j=1}^{N}j_{2,ij}\Psi_i^{\dagger}\Psi_j\
\label{Hj2}
\end{equation}
where again the couplings $j_{2,ij}$ are random numbers drawn from a Gaussian distribution. Henceforth this system will be called $(q=2,4)$ SYK model. There has been claims that this system also undergoes chaotic-integrable phase transition. But from explicit calculation of the partition function, we found that there is no phase transition. The chaotic dynamics is increasingly suppressed when the $q=2$ interaction strength is increased but the system is never completely integrable. This agrees with the result of high precision calculation of the Lyapunov exponent of this system \cite{Tousik2020}. The Lyapunov exponent never goes to zero completely. This also agrees with the result from non-equilibrium dynamics. It has been found that the $q=2$ interaction slows down the thermalization process, but the system ultimately thermalizes even when $q=2$ interaction strength is very large \cite{Bhattacharya:2018fkq}. Despite this, we expect that turning on chemical potential would still give rise to a phase transition even in the presence of $(q=2)$ interaction.

The technical advancement in this work is the use of a novel technique which immensely speed up the numerical calculations and more importantly makes it possible to explicitly calculate the partition functions. The technique is to use different UV limits for Euclidean time domain and frequency domain. The standard prescription of numerical calculations in thermal (fermionic) quantum systems at inverse temperature $\beta$ is to use time intervals $\beta/L$ and frequency range $\left\{-\frac{2\pi}{\beta}\frac{L'-1}{2},\dots,-\frac{2\pi}{\beta}\frac{1}{2},\frac{2\pi}{\beta}\frac{1}{2},\dots,\frac{2\pi}{\beta}\frac{L'-1}{2}\right\}$ where $L'=L$. But in our case, we will use different $L$ and $L'$. This relies on the fact that the UV limit of our theory is a free (solvable) theory. This speeds up the process of solving the SD equations by a factor of 10 and also makes it possible to explicitly calculate the determinant in the expression of partition function by reducing memory requirement by a factor of 100 and speeding up matrix decomposition by a factor of 1000. Otherwise an extremely large amount of computer memory and lengthy computer time of high performance computing facilities would have been required. Details on this aspect can be found in section \ref{numtech}.

In previous works \cite{Maldacena:2016hyu,Azeyanagi:2017drg,Maldacena:2018lmt}, the calculation of partition functions exploits the fact that in SYK models with a single interaction term (like $q=4$ SYK model) there are only two dimensionful parameters in the theory - the inverse temperature $\beta$ and the coupling strength $J_q$. This technique is still applicable in the presence of conserved charges. But this technique fails in models like $(q=2,4)$ SYK model. We believe that this is one reason why the partition function of $(q=2,4)$ SYK model has not been calculated so far. Now with our new technique, it is possible to explicitly calculate the partition function of this model.

In real time dynamics, mass and chemical potential are different. Chemical potential is manifested in the state while mass term is a part of the Hamiltonian. We work out the differences for simple free theories in Appendix \ref{massversuschempot}. So, for the system with hamiltonian $H_{SYK}$ we consider mixed states with the following probability densities.
\begin{gather}
\rho_1(\beta)=e^{-\beta H_{SYK}}\,, \qquad \rho_2(\beta,\eta)=e^{-\beta(H_{SYK}+\eta Q)}\
\end{gather}
For the system with Hamiltonian $\tilde{H}_{SYK}$, we consider the states
\begin{gather}
\tilde{\rho}_1(\beta)=e^{-\beta \tilde{H}_{SYK}}\,, \qquad \tilde{\rho}_2(\beta,\eta)=e^{-\beta(\tilde{H}_{SYK}+\eta Q)}\
\end{gather}
The effective chemical potential of $\tilde{\rho}_1(\beta)$ is $\mu$ and of $\tilde{\rho}_2(\beta)$ is $\mu+\eta$. An interesting case is when we take $\eta=-\mu$ in $\tilde{\rho}_2(\beta,\eta)$. The probability density is effectively $\rho_1(\beta)$, but the time evolution operator is still $\tilde{H}_{SYK}$.

There are many conjectured diagnostics of chaos. One of the most popular test is the comparison of the energy spectrum with the eigenvalue spectrum of random matrix \cite{Bohigas:1983er}. Another popular test is to examine spectral form factor \cite{Papadodimas:2015xma,Cotler:2016fpe}. In this paper, we will calculate Out-of-Time-Ordered corellators (OTOC). OTOC of chaotic systems like SYK model grows exponentially \cite{Maldacena:2015waa}
\begin{eqnarray}
C(t)&=&\text{Tr}\langle e^{-\beta H_{SYK}/4}\Psi_i^{\dagger}(t)e^{-\beta H_{SYK}/4}\Psi_j^{\dagger}(0)e^{-\beta H_{SYK}/4}\Psi_i(t)e^{-\beta H_{SYK}/4}\Psi_j(0)\rangle\nonumber\\
&=& f_0-\frac{f_1}{N}\, e^{\lambda_L t}+\mathcal{O}(N^{-2})\
\end{eqnarray}
where $N$ is the number of degrees of freedom in the system. It has been conjectured that the Lyapunov exponent $\lambda_L$ of a quantum system has a upper bound.
\begin{equation}
\lambda_L \leq \frac{2\pi}{\beta}\,,\qquad \lambda^*_L=\frac{\lambda_L \beta}{2\pi}\leq 1\
\label{chaosbound}
\end{equation}
Interestingly there has been exceptional cases where this bound has been found to be violated \cite{Poojary:2018esz,David:2019bmi,Halder:2019ric}. The most interesting case is the bulk calculation of the BTZ black hole background with non-zero angular momentum \cite{Poojary:2018esz}. For this particular case, the violation of the above bound has been attributed to the fact that left-moving and right-moving degrees of freedom in the boundary theory have different effective temperatures $\beta_{\pm}=(\beta \pm J)$ \cite{Jahnke:2019gxr}. Interestingly, it has also been claimed that the growth rate of the OTOC oscillates with the average growth rate given by $2\pi/\beta$ \cite{Mezei:2019dfv}. Lyapunov exponent in charged SYK model has also been calculated in the large $q$ limit where the chemical potential is absorbed by redefining an effective coupling strength \cite{Bhattacharya:2017vaz}. In this case, the Lyapunov exponent is found to be greatly suppressed by the chemical potential. But the upper bound is still given by the above equation (\ref{chaosbound}). This can be easily shown in our case by using the two Baker–Campbell–Hausdorff (BCH) relations.
\begin{eqnarray}
\label{BCH1}
\exp^{\kappa Q} \Psi^{\dagger}&=&\exp^{\kappa} \Psi^{\dagger}\exp^{\kappa Q}\\
\exp^{\kappa Q} \Psi &=& \exp^{-\kappa} \Psi\exp^{\kappa Q}\
\label{BCH2}
\end{eqnarray}
Using this relations, the OTOC of the system with Hamiltonian $H_{SYK}$ in the thermal state with chemical potential $\rho_2(\beta,\eta)$ is
\begin{eqnarray}
\text{OTOC}&=&\text{Tr}\left(e^{-\beta(H+\eta Q)/4}\,\psi_i^{\dagger}(t)\,e^{-\beta(H+\eta Q)/4}\,\psi_j^{\dagger}(0)\,e^{-\beta(H+\eta Q)/4}\,\psi_i(t)\,e^{-\beta(H+\eta Q)/4}\,\psi_j(0)\right)\nonumber\\
&=&\text{Tr}\left(e^{-\beta(H+\eta Q)/4}\,e^{itH}\psi_i^{\dagger}(0)e^{-itH}\,e^{-\beta(H+\eta Q)/4}\,\psi_j^{\dagger}(0)\right.\nonumber\\
&& \qquad \qquad \times \,\left. e^{-\beta(H+\eta Q)/4}\,e^{itH}\psi_i(0)e^{-itH}\,e^{-\beta(H+\eta Q)/4}\,\psi_j(0)\right)\nonumber\\
&=&\text{Tr}\left(e^{-\beta \tilde{H}(\eta)/4}e^{-it \eta Q}\,e^{it\tilde{H}(\eta)}\psi_i^{\dagger}(0)e^{-it\tilde{H}(\eta)}e^{it \eta Q}\,e^{-\beta \tilde{H}(\eta)/4}\,\psi_j^{\dagger}(0)\right.\nonumber\\
&& \qquad \qquad \times \,\left. e^{-\beta\tilde{H}(\eta)/4}\,e^{-it \eta Q}e^{it\tilde{H}(\eta)}\psi_i(0)e^{-it\tilde{H}(\eta)}e^{it \eta Q}\,e^{-\beta \tilde{H}(\eta)/4}\,\psi_j(0)\right)\nonumber\\
&=& \text{OTOC for} \,\, \tilde{\rho}_1(\beta)\,\, \text{state with Hamiltonian}\,\, \tilde{H}_{SYK}(\mu=\eta)\
\label{OTOCequiv}
\end{eqnarray}
Using the same line of argument provided in \cite{Maldacena:2015waa}, the upper bound is still $2\pi/\beta$. Similar general argument has also been given in \cite{Halder:2019ric}.

The suppression of Lyapunov exponent at finite temperature is not surprising. It can be easily seen in the extreme case of $\eta\to\infty$ with $\beta$ finite, the Lyapunov exponent is zero. The state in this limit is all empty state $|0\rangle$.
\begin{eqnarray}
\langle 0|e^{-\beta (H_{SYK}+\eta Q)/4}\Psi_i^{\dagger}(t)e^{-\beta (H_{SYK}+\eta Q)/4}\Psi_j^{\dagger}(0)e^{-\beta (H_{SYK}+\eta Q)/4}\Psi_i(t)e^{-\beta (H_{SYK}+\eta Q)/4}\Psi_j(0)|0\rangle
= 0\nonumber\\
\label{etaextreme1}
\end{eqnarray}
Similarly with $\eta\to-\infty$, the state is all filled state $|1\rangle$ and again the Lyapunov exponent is zero. Note that $|1\rangle$ and $|0\rangle$ are eigenstates of $Q$ so they are also eigenstates of $H_{SYK}$.

The suppression of the Lyapunov exponent is solely due to the chemical potential.    If one considers the Hamiltonian $\tilde{H}_{SYK}$, the mass term does not affect the Lyapunov exponent, it only depends on the effective chemical potential. The simplest case is when we consider $\eta=-\mu$ in the state $\tilde{\rho}_2(\beta,\eta)$. Just like the derivation of equation (\ref{OTOCequiv}), we can show that the OTOC in this particular state is equal to the OTOC of the state $\rho_1(\beta)$ with Hamiltonian $H_{SYK}$.

Another viewpoint of OTOC is operator scrambling \cite{Roberts:2014isa}. It measures the rate of growth of an operator with time evolution. From this viewpoint the suppression implies that the state with chemical potential picks out `operator strings' with slower growth rate. The fastest growing operator strings are killed by the state. Their expectation values are negligible. Operator growth in SYK model has been worked out in \cite{Roberts:2018mnp, Qi:2018bje}.

An interesting consequence of the BCH relations (\ref{BCH1},\ref{BCH2}) is that mass quench\footnote{Examples of mass quench in simple 2-D theories can be found in \cite{Mandal:2015kxi}.} is trivial in 1-D quantum systems including SYK model. For example consider a mass quench, starting from a thermal state of a massive theory, one turns off the mass term (at any rate). The final state is a thermal state of the same temperature but with chemical potential turned on. The chemical potential being equal to the initial mass. The Green's functions after the quench is simply given by the relation (\ref{Grelation}). The equilibration process is instantaneous. As soon as both the time arguments passed the quench region, the Green's functions reach their final values. Similar BCH relations exist for bosonic systems also. The mass or the charge cannot be negative in bosonic systems.

\vspace{0.5cm}

The main results of this work are as follows:
\begin{enumerate}
\item From explicit calculation of the partition function, we show the phase transition in $(q=4)$ SYK model with chemical potential. We find that large temperature jump during the cooling process induce the transition from the chaotic state to the integrable state.
\item There is no sharp phase transition in $(q=2,4)$ SYK model.
\item We also observe the phase transition due to the presence of charge in real time solutions of the $(q=4)$ SYK model. The Lyapunov exponent is non-zero in the chaotic phase and it is effectively zero in the integrable phase. The Lyapunov exponent sharply goes to zero at the transition point. The presence of the chemical potential suppresses the Lyapunov exponent exponentially.
\end{enumerate}

\vspace{0.5cm}

One of the most important consequences of these results is that thermalization could be state-dependent. If one prepares the system in the chaotic state, the system would thermalize after a time-dependent perturbation. But instead if the same system (with the exactly same Hamiltonian) was prepared in the integrable phase, it would not thermalize. This is the subject of our ongoing work \cite{Tousik2020}. Interesting work in this direction has also been carried out in \cite{Haldar:2019slc}.\footnote{But unlike our present model, the two phases exist in separate parameter ranges of the interaction strengths for the model considered in \cite{Haldar:2019slc}.} We also would like to point out that interesting gravity configurations with no black hole formation have been found \cite{Craps:2014eba,lindgren2019black}. More details can be found in section \ref{cnd}. The sharp decay of the Lyapunov exponent at the transition point also has interesting analog in gravity \cite{Ageev:2018msv}.

There are two paradigms in which thermalization of a chaotic system is slowed down (or stopped). One is the existence of quantum scars \cite{PhysRevLett.53.1515} of chaotic hamiltonians. Quantum scar is a topic of intense research in theoretical and experimental condensed-matter physics \cite{Bernien_2017,Turner2018WeakEB,PhysRevB.98.155134,Ho_2019,Khemani_2019,Lin_2019,mukherjee2019collapse}. Quantum scars are special states in chaotic systems which do not thermalize and which violate the eigenstate thermalization hypothesis (ETH). They are usually considered in systems without disorder. The other paradigm is many body localization (MBL) \cite{Pal_2010,Nandkishore_2015,Alet_2018,Abanin_2019} where a solvable (free) random interaction slows down the thermalization process. MBL is also presently a topic of intense research interest. It is also claimed that MBL is a dynamical phase transition. MBL is also usually considered in chaotic systems without disorders in which the random interaction is introduced.

The integrable states that we found in this work are similar to quantum scars. Because these are very special states which would not thermalize although the Hamiltonian of the system is highly chaotic. But it should be noted that quantum scars are pure states and have many other properties like weak entanglement, almost closed Hilbert subspace, etc.\footnote{The author thanks Krishnendu Sengupta for helpful discussion on quantum scars.} On the other hand, the suppression of chaos due to the presence $(q=2)$ SYK term is more akin to MBL. (q=2) SYK term is a solvable interaction. One can diagonalize this term which by itself gives non-interacting free fermions of random mass. So besides the fact that the chaotic part of the Hamiltonian ($q=4$ interaction) is disordered, the suppression of chaos in $(q=2,4)$ SYK model is very much like MBL. Our result of $(q=2,4)$ SYK model suggests that MBL is not a sharp transition.

The outline of this paper is as follows: In section \ref{csykim}, we introduce complex SYK model in imaginary time formalism and reproduce the Schwinger-Dyson equations of the theory. In section \ref{numtech}, we elaborate on the numerical recipe used to solve the SD equations. In section \ref{phasetrans},  we calculate the partition function of $(q=4)$ SYK model with chemical potential and show the phase transition. In section \ref{sec:q24}, we calculate the partition of $(q=2,4)$ SYK model. We show that there is no sharp phase transition for this model. In section \ref{csykrt}, we solve the SD equation in real time formalism. In section \ref{sec:lyap}, we calculate OTOC and the associated Lyapunov exponents. Section \ref{cnd} consists of conclusions. The appendix consists of a collection of general results we have used in the main text. Appendix \ref{conventions} consists of the conventions we have used in the paper. In Appendix \ref{massversuschempot}, we differentiate mass and chemical potential. Fluctuation-Dissipation theorem in the presence of chemical potential is derived in Appendix \ref{fdchem}. Appendix \ref{partfunccalc} consists of details and subtleties involved in numerical calculation of partition function of a fermionic theory.

\section{Complex SYK model in imaginary time formalism}
\label{csykim}
The (grand) partition function is
\begin{equation}
Z(\beta,\mu)=\text{Tr} \, e^{-\beta (H_{SYK}+\mu Q)}=\text{Tr} \, e^{-\beta \tilde{H}_{SYK}}\
\end{equation}
As pointed out above, mass and chemical potential are equivalent in imaginary time formalism. We will consider the Hamiltonian with both $q=2$ and $q=4$ interactions. The Hamiltonian and the charge are
\begin{equation}
H_{SYK}=\sum_{i,j=1}^Nj_{2,ij}\Psi^{\dagger}_i\Psi_j+\sum_{i,j,k,l=1}^{N} j_{4,ij;kl} \, \Psi^{\dagger}_i \Psi^{\dagger}_j \Psi_k \Psi_l\,\qquad Q=\sum_{i=1}^{N} \Psi_i^{\dagger}\Psi_i\
\label{Hsyk}
\end{equation}

In path integral language, the partition function in terms of grassmann fields $\psi_i(\tau)$ is
\begin{equation}
Z=\int \mathcal{D}\psi^{\dagger}_i \mathcal{D}\psi_i \,\,
\text{exp}\left[-\int_0^{\beta}d\tau \, \left\{\psi^{\dagger}_i\partial_{\tau}\psi_i+H(\psi^{\dagger}_i,\psi_i)+\mu Q(\psi^{\dagger}_i,\psi_i)\right\}\right]
\end{equation}

The symmetries of the couplings are
\begin{gather}
j_{2,ij}^*=j_{2,ji} \qquad j_{4,ij;kl}=j^*_{4,kl;ij}\\
j_{4,ij;kl}=-j_{4,ji;kl},\qquad j_{4,ij;kl}=-j_{4,ij,lk}\
\label{Jsym}
\end{gather}
These symmetries ensure that the Hamiltonian is hermitian. $j_{2,ij}, j_{4,ij;kl}$ are disordered couplings with a Gaussian distribution. Instead of the range of indices in (\ref{Hsyk}), below we will consider $i<j$ and $k<l$ for $j_{4,ij;kl}$, while $j$ and $l$ runs from 1 to $N$. With this convention, we separate the real and imaginary parts of $j_{4,ij;kl}$.
\begin{gather}
j_{2,ij}=j_{2R,ij}+ij_{2I,ij},\qquad \langle j_{2R,ij}\rangle =\langle j_{2I,ij}\rangle=0\\
j_{4,ij;kl}=j_{4R,ij;kl}+ij_{4I,ij;kl},\qquad \langle j_{4R,ij;kl}\rangle =\langle j_{4I,ij;kl}\rangle=0\\
\langle j_{2R,ij}^2\rangle=\langle j_{2I,ij}^2\rangle=J^2_2, \qquad \langle j_{4R,ij;kl}^2\rangle=\langle j_{4I,ij;kl}^2\rangle=J^2_4\
\end{gather}
The exact distributions are
\begin{eqnarray}
\mathcal{P}_{2R}\left(j_{2R,ij}\right)&=&\left(\sqrt{\frac{N}{2J_2^2\pi}}\right)^{N(N-1)}\;\exp\left(-\frac{N}{2J_2^2}\sum_{i<j}j_{2R,ij}^2\right)\\
\mathcal{P}_{2I}\left(j_{2I,ij}\right)&=&\left(\sqrt{\frac{N}{2J_2^2\pi}}\right)^{N(N-1)}\;\exp\left(-\frac{N^3}{2J_2^2}\sum_{i<j}j_{2I,ij}^2\right)\\
\mathcal{P}_{4R}\left(j_{4R,ij;kl}\right)&=&\left(\sqrt{\frac{N^3}{J_4^2\pi}}\right)^{N(N-1)}\;\exp\left(-\frac{N^3}{J_4^2}\sum_{i<j,k<l}j_{4R,ij;kl}^2\right)\\
\mathcal{P}_{4I}\left(j_{4I,ij;kl}\right)&=&\left(\sqrt{\frac{N^3}{J_4^2\pi}}\right)^{N(N-1)}\;\exp\left(-\frac{N^3}{J_4^2}\sum_{i<j,k<l}j_{4I,ij;kl}^2\right)\
\end{eqnarray}
We work with quenched averaging of the disordered couplings where we perform averaging at the level of the partition function.
\begin{equation}
Z=\int \mathcal{D}\psi^{\dagger} \mathcal{D}\psi \int \mathcal{D}j_{4R,ij;kl}\mathcal{D}j_{4I,ij;kl} \mathcal{P}_{2R}\left(j_{2R,ij}\right) \, \mathcal{P}_{2I}\left(j_{2I,ij}\right) \, \mathcal{P}_{4R}\left(j_{4R,ij;kl}\right) \, \mathcal{P}_{4I}\left(j_{4I,ij;kl}\right) \, e^{-S[\psi^{\dagger},\psi]}\
\end{equation}
The action is
\begin{eqnarray}
S[\psi^{\dagger},\psi]&=&\int_{\mathcal{C}} d\tau \left\{\;\sum_i\psi^{\dagger}_i\partial_{\tau} \psi_i +\mu \sum_i\psi^{\dagger}_i\psi_i+\sum_{i<j}\left[j_{2R,ij}\left(\psi^{\dagger}_i\psi_j+\psi^{\dagger}_j\psi_i\right)+ij_{2I,ij}\left(\psi^{\dagger}_i\psi_j-\psi^{\dagger}_j\psi_i\right)\right]\right.\nonumber\\
&& \left.\quad + \sum_{i<k,j<l}\left[j_{4R,ij;kl}\,\left(\psi^{\dagger}_i \psi^{\dagger}_j \psi_k \psi_l + \psi^{\dagger}_k \psi^{\dagger}_l \psi_i \psi_j\right)+i\,j_{4I,ij;kl}\,\left(\psi^{\dagger}_i \psi^{\dagger}_j \psi_k \psi_l - \psi^{\dagger}_k \psi^{\dagger}_l \psi_i \psi_j\right)\right]\right\}\nonumber\\
\end{eqnarray}
Performing the Gaussian integrals of $j_{2R,ij}$, $j_{2I,ij}$, $j_{4R,ij;kl}$ and $j_{4I,ij;kl}$, we get
\begin{eqnarray}
Z&=&\int \mathcal{D}\psi^{\dagger} \mathcal{D}\psi \; \exp\left[-\int_{\mathcal{C}} d\tau \sum_i\psi^{\dagger}_i(\partial_\tau+\mu) \psi_i+\frac{J_2}{N}\int_{\mathcal{C}} dt_1 dt_2 \sum_{i<j}\left\{P_{1,ij}-P_{2,ij}\right\}\right.\nonumber\\
&&\qquad \left. +\frac{2J_{4}^2}{4N^3}\int_{\mathcal{C}} dt_1 dt_2 \sum_{i<k,j<l}\left\{P_{3,ijkl}(t_1,t_2) - P_{4,ijkl}(t_1,t_2)\right\}\right]\nonumber\\
\end{eqnarray}
where
\begin{eqnarray}
P_{1,ij}&=&\left(\psi^{\dagger}_i(\tau_1)\psi_j(\tau_1)+\psi^{\dagger}_j(\tau_1)\psi_i(\tau_1)\right)\left(\psi^{\dagger}_i(\tau_2)\psi_j(\tau_2)+\psi^{\dagger}_j(\tau_2)\psi_i(\tau_2)\right)\nonumber\\
P_{2,ij}&=&\left(\psi^{\dagger}_i(\tau_1)\psi_j(\tau_1)-\psi^{\dagger}_j(\tau_1)\psi_i(\tau_1)\right)\left(\psi^{\dagger}_i(\tau_2)\psi_j(\tau_2)-\psi^{\dagger}_j(\tau_2)\psi_i(\tau_2)\right)\nonumber\\
P_{3,ijkl}&=&\left(\psi^{\dagger}_i(\tau_1)\psi^{\dagger}_j(\tau_1)\psi_k(\tau_1)\psi_l(\tau_1)+\psi^{\dagger}_k(\tau_1) \psi^{\dagger}_l(\tau_1) \psi_i(\tau_1) \psi_j(\tau_1)\right)\nonumber\\
&&\qquad \times\left(\psi^{\dagger}_i(\tau_2)\psi^{\dagger}_j(\tau_2)\psi_k(\tau_2)\psi_l(\tau_2)+\psi^{\dagger}_k(\tau_2) \psi^{\dagger}_l(\tau_2) \psi_i(\tau_2) \psi_j(\tau_2)\right)\nonumber\\
P_{4,ijkl}&=&\left(\psi^{\dagger}_i(\tau_1)\psi^{\dagger}_j(\tau_1)\psi_k(\tau_1)\psi_l(\tau_1)-\psi^{\dagger}_k(\tau_1) \psi^{\dagger}_l(\tau_1) \psi_i(\tau_1) \psi_j(\tau_1)\right)\nonumber\\
&&\qquad \times \left(\psi^{\dagger}_i(\tau_2)\psi^{\dagger}_j(\tau_2)\psi_k(\tau_2)\psi_l(\tau_2)-\psi^{\dagger}_k(\tau_2) \psi^{\dagger}_l(\tau_2) \psi_i(\tau_2) \psi_j(\tau_2)\right)\nonumber\
\end{eqnarray}
Terms like $\psi^{\dagger}_i(\tau_1)\psi_j(\tau_1)\psi^{\dagger}_i(\tau_2)\psi_j(\tau_2)$, $\psi^{\dagger}_i(\tau_1)\psi^{\dagger}_j(\tau_1)\psi_k(\tau_1)\psi_l(\tau_1)\psi^{\dagger}_i(\tau_2)\psi^{\dagger}_j(\tau_2)\psi_k(\tau_2)\psi_l(\tau_2)$, which are products of $\psi^{\dagger}_i(\tau)\psi^{\dagger}_i(\tau')$ and $\psi_i(\tau)\psi_i(\tau')$, get cancelled due to the symmetries of $j_{2,ij}$ and $j_{4,ij;kl}$, and the Gaussian integrations. Finally we have,
\begin{eqnarray}
Z&=&\int \mathcal{D}\psi^{\dagger} \mathcal{D}\psi \; \exp\left[-\int_{\mathcal{C}} d\tau \sum_i\psi^{\dagger}_i(\partial_{\tau}+\mu) \psi_i - \frac{J^2}{N}\int_{\mathcal{C}} d\tau_1 d\tau_2 \sum_{i,j=1}^N \psi_i(\tau_2)\psi^{\dagger}_i(\tau_1)\psi_j(\tau_1)\psi^{\dagger}_j(\tau_2)\right.\nonumber\\
&& \left.+\frac{J_{4}^2}{2N^3}\int_{\mathcal{C}} d\tau_1 d\tau_2 \sum_{i,k,j,l=1}^N \; \psi_i(\tau_2)\psi^{\dagger}_i(\tau_1)\psi_j(\tau_2)\psi^{\dagger}_j(\tau_1)\psi_k(\tau_1)\psi^{\dagger}_k(\tau_2)\psi_l(\tau_1)\psi^{\dagger}_l(\tau_2)\right]\
\end{eqnarray}
In large N limit, only melonic diagrams dominate so we will enforce \begin{equation}
\mathcal{G}(\tau_1,\tau_2)=\frac{1}{N}\sum_i\langle\mathcal{T}_{\tau}(\psi_i(\tau_1)\psi^{\dagger}_i(\tau_2))\rangle\
\label{therm_prop}
\end{equation}
using a Lagrange multiplier $\Sigma(\tau_1,\tau_2)$ which turns out to be the self energy. So, the partition function is
\begin{eqnarray}
Z
&=&\int \mathcal{D}\psi^{\dagger}_i \mathcal{D}\psi_i \mathcal{D}G \mathcal{D}\Sigma \,\,
\text{exp}\left[-\int_0^{\beta}d\tau_1 d\tau_2 \, \sum_i\psi^{\dagger}_i(\tau_1)\left\{\delta(\tau_1-\tau_2)(\partial_{\tau_2}+\mu)+\Sigma(\tau_1,\tau_2)\right\}\psi_i(\tau_2)\right.\nonumber\\
&&\quad \left.+\,N\int_0^{\beta}d\tau_1 d\tau_2 \left\{-\frac{J^2_2}{2}\, \mathcal{G}(\tau_1,\tau_2)\mathcal{G}(\tau_2,\tau_1)+\frac{J_4^2}{4}\, \mathcal{G}(\tau_1,\tau_2)^2\mathcal{G}(\tau_2,\tau_1)^2-\Sigma(\tau_1,\tau_2)\mathcal{G}(\tau_2,\tau_1)\right\}\right]\nonumber\
\label{part_func_first}
\end{eqnarray}
Performing the $\psi^{\dagger}_i$ and $\psi_i$ grassmanian integrals, we get
\begin{eqnarray}
Z&=&Det\left[\delta(\tau_1-\tau_2)(\partial_{\tau_2}+\mu)+\Sigma(\tau_1,\tau_2)\right]^N\nonumber\\
&& \times\,\text{exp}\left[N\int_0^{\beta}d\tau_1 d\tau_2 \left\{-\frac{J^2_2}{2}\, \mathcal{G}(\tau_1,\tau_2)\mathcal{G}(\tau_2,\tau_1)+\frac{J_4^2}{4}\, \mathcal{G}(\tau_1,\tau_2)^2\mathcal{G}(\tau_2,\tau_1)^2-\Sigma(\tau_1,\tau_2)\mathcal{G}(\tau_2,\tau_1)\right\}\right]\nonumber\\
\label{partfunc}
\end{eqnarray}
Using the convention defined in (\ref{psimode}) for the Fourier transforms of the grassmann variables, the equations of motion are
\begin{gather}
\label{SD1}
\Sigma(\tau_1,\tau_2)=-J^2_2\,\mathcal{G}(\tau_1,\tau_2)+J_4^2\,\mathcal{G}(\tau_1,\tau_2)^2\mathcal{G}(\tau_2,\tau_1)\\
\mathcal{G}(\omega_n)=\frac{1}{-i\omega_n+\mu+\Sigma(\omega_n)}\
\label{SD2}
\end{gather}
These are the Schwinger-Dyson(SD) equation of complex SYK model with (q=2) and (q=4) interactions and non-zero chemical potential.

\subsection{Numerical method}
\label{numtech}
We solved the SD equations (\ref{SD1}, \ref{SD2}) numerically. The two equations form a closed iterative loop. The loop is executed until the desired convergence is achieved. An approximated initial values of $\mathcal{G}(\omega_n)$ are used to start the iterations. We used the propagators of the exactly solvable $q=2$ SYK model as the initial values. The thermal propagator for (q=2) SYK model is
\begin{equation}
\mathcal{G}(\omega_n)=\frac{2i}{\omega_n+\,\text{sign}(\omega_n)\sqrt{\omega_n^2+4J_2^2}}\
\label{iniGim}
\end{equation}
This initial data is used to solve the SD equations at a relatively high temperature. After this the new solution is used to solve the SD equations for lower temperatures. So, we can gradually cool down or heat up the system.\footnote{Note that we are not considering coupling a heat bath to the system like the set-up considered in \cite{Maldacena:2019ufo}. We only meant considering an initial field configuration and taking it to another field configuration using the numerical steps mentioned.}

The test for convergence is done by calculating
\begin{equation}
\Delta\mathcal{G}=\sum_{\omega_n}\left|\mathcal{G}(\omega_n)-\mathcal{G}_{prev}(\omega_n)\right|\
\label{delGim}
\end{equation}
where $\mathcal{G}_{prev}(\omega_n)$ is the Green's function in the previous iteration. The iteration is stopped when $\Delta\mathcal{G}$ is smaller than a preset tolerance limit. One of the crucial numerical technique used to achieve convergence is the weighted iteration pointed out in \cite{Maldacena:2016upp}. We take half weight where we take half of the previous value of $\mathcal{G}(\omega_n)$ and updated the other half using (\ref{SD2}).

We choose the imaginary time interval $\{-\beta/2,\beta/2\}$ for performing the iterative loops instead of the usual $(0,\beta)$. But once the desired convergence is achieved, we can perform a Fourier transform and calculate the propagator at any range of the imaginary time. The above imaginary time interval is discretized into $L$ points with the interval between each adjacent points being $\beta/L$. We mostly take $L=10000$. We found that taking any lesser number of points introduce large errors in grand potential calculation. In the frequency space, we used the range
\begin{equation}
\left\{-\frac{2\pi}{\beta}\frac{L'-1}{2},\dots,-\frac{2\pi}{\beta}\frac{1}{2},\frac{2\pi}{\beta}\frac{1}{2},\dots,\frac{2\pi}{\beta}\frac{L'-1}{2}\right\}\nonumber\
\end{equation}
with interval size $2\pi/\beta$. The standard prescription of thermal quantum theory is to take $L=L'$. But here we take $L'$ much larger than $L$. We mostly take $L'=10^5$ and larger is better. Taking this higher UV cut-off produces more accurate values of $\mathcal{G}(\tau_1,\tau_2)|_{\tau_1\to\tau_2^-}$ and $\mathcal{G}(\tau_1,\tau_2)|_{\tau_1\to\tau_2^+}$ which in turn gives more accurate value of grand potential.

The justification for the asymmetric UV limits $L$ and $L'$ is that at all other points $L=L'=10^4$ would work extremely well except for few points in the neighbourhood of $\tau_1=\tau_2$. We can see this in the thermal propagator plots in Figure \ref{fig:therm_prop}. For any sensible fermionic thermal propagator, there is a sharp discontinuity at $\tau_1=\tau_2$. This comes from the definition of the thermal propagator (\ref{therm_prop}) and the fermionic commutation relations. So, the larger value of $L'$ captures this special neighbourhood very well. Technically, this appears to be incorrect. But what is really happening is that for large numerical values of $\omega_n$, we are relying on the approximation
\begin{equation}
\mathcal{G}(\omega_n)=\frac{1}{i\omega_n+\mu+\Sigma(\omega_n)} \approx \frac{1}{i\omega_n+\mu}\
\label{SDapprox}
\end{equation}
because $\Sigma(\omega_n)$ is numerically very small compare to $\omega_n$. The above approximation is the propagator of the free UV theory. So, we are using the correct solution in the high energy limit with which we can take the UV limit as large as possible irrespective of the time domain discretization. This technique can be generalized and applied to any theory with a solvable UV limit. In other word, it is almost as if we are using the correct $L=L'=10^5$ UV limit but we sample only $L=10^4$ points in time domain for the solution $\mathcal{G}(\tau_1,\tau_2)$.

This strategy have two-fold benefits:
\begin{enumerate}
\item First, it makes it possible to solve the SD equations even in a modern personal computer for the parameter ranges that we are studying. We used a desktop computer with Intel i7-7700 processor and 8 Gigabytes of memory. One iteration to solve the SD equations using all available cores in the computer processor took around 37 seconds with the above values of $L=10^4$ and $L'=10^5$. While using $L=L'=10^5$, one iteration took around 380 seconds. So, the strategy to use different values of $L$ and $L'$ immensely speed up the numerical calculation of the solutions of the SD equations.
\item Secondly and more crucially, it would have been impossible to directly calculate the determinant in (\ref{part_func_first}) if one uses $L=L'=10^5$ due to memory constraint. But with $L=10^4$, the memory required is around 800 Megabytes. Moreover, with the desktop computer mentioned above, it took around 10 seconds to decompose the $10^4\,\times\,10^4$ real matrix with double precision entries. With $L=10^5$, around 80 Gigabytes of memory would have been required. The computer time for matrix decomposition goes as $\mathcal{O}(L^3)$.\footnote{The actual implementation of the matrix decomposition that we used is little faster than $\mathcal{O}(L^3)$.} So, it would have taken $10^4$ seconds!
\end{enumerate}

We have checked that the $G(\omega_n)$ solved using $L=10^4, L'=10^5$ also solve the SD equation equation when we take $L=L'=10^5$. But we could not calculate the partition function due to the memory constraint. Instead we have checked that the partition function values calculated using different values of $L=10^4$ and $2\,\times 10^4$ match and they match very well.

\section{Phase transition in $(q=4)$ model with chemical potential}
\label{phasetrans}
The grand potential (per fermion) is defined as
\begin{equation}
\Omega(\beta,\mu)=-\frac{1}{\beta N}\log\left[Z(\beta,\mu)\right]\
\label{gpdef}
\end{equation}
In this section we will consider the theory without (q=2) SYK interaction (or $J_2=0$). The expression for the partition function is
\begin{eqnarray}
\label{partfunc2}
Z&=&Det\left[\delta(\tau_1-\tau_2)(\partial_{\tau_2}+\mu)+\Sigma(\tau_1,\tau_2)\right]^N\;e^{\left[-\frac{3N\beta}{4}\int_0^{\beta}d\tau_1 d\tau_2 \; \Sigma(\tau_1,\tau_2)\mathcal{G}(\tau_2,\tau_1)\right]}\\
\text{where} && \qquad \Sigma(\tau_1,\tau_2)=J_4^2\,\mathcal{G}(\tau_1,\tau_2)^2\mathcal{G}(\tau_2,\tau_1)\nonumber\
\end{eqnarray}
The grand potential is calculated using the above two equations after solving the SD equations. But note that straight-forward use of the above equations does not give the correct grand potential or the correct partition function. Even for free theory, $Det\left[\partial_{\tau}+\mu\right]$ does not numerically reproduce the free fermion partition function. We used the numerical recipe in Appendix \ref{partfunccalc}.

\begin{figure}[h]
\centering
\begin{subfigure}{.5\textwidth}
  \centering
  \includegraphics[width=.9\linewidth]{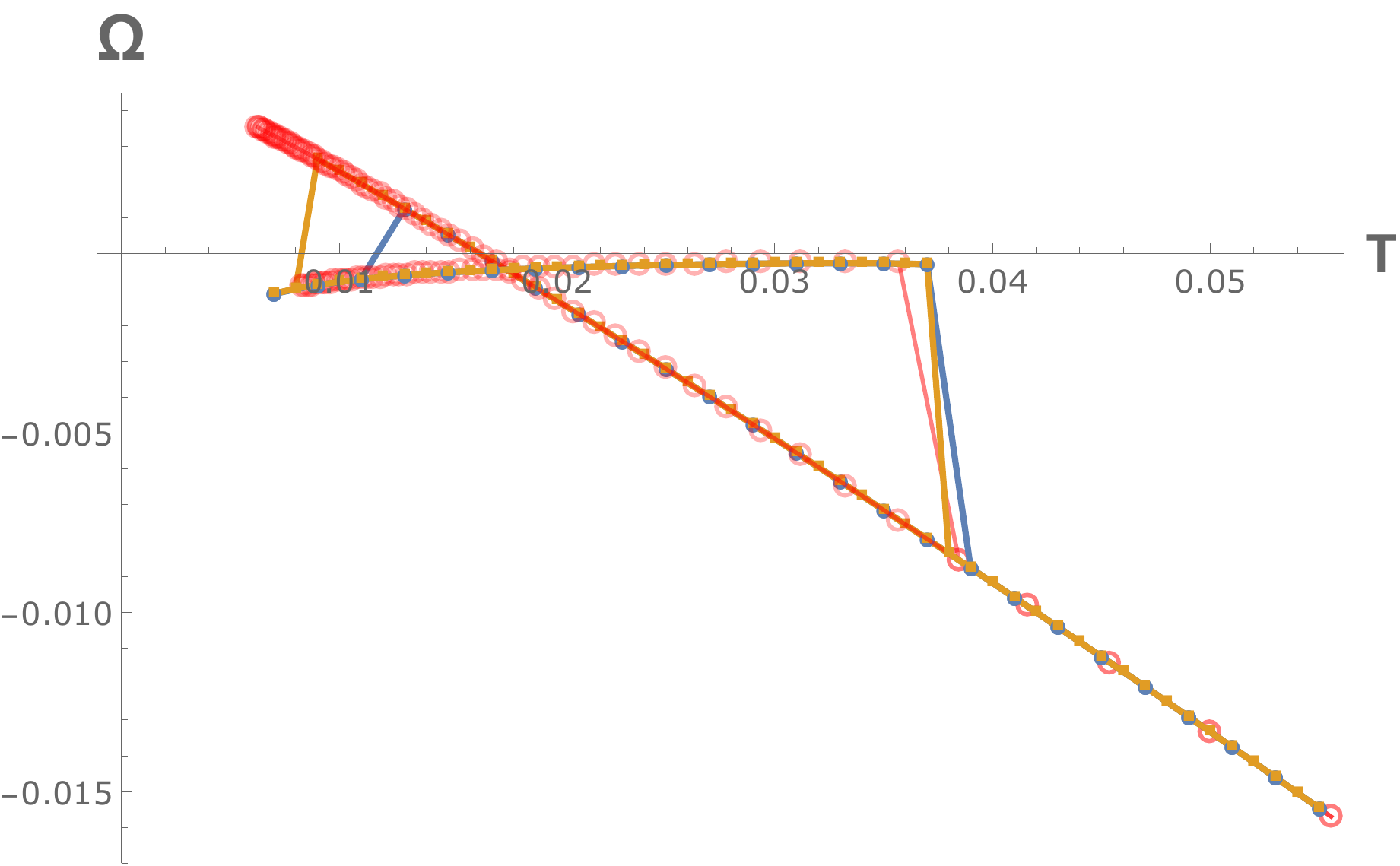}
  \caption{$\mu=0.24$}
  \label{fig:phasetransition_mu024}
\end{subfigure}%
\begin{subfigure}{.5\textwidth}
  \centering
  \includegraphics[width=.9\linewidth]{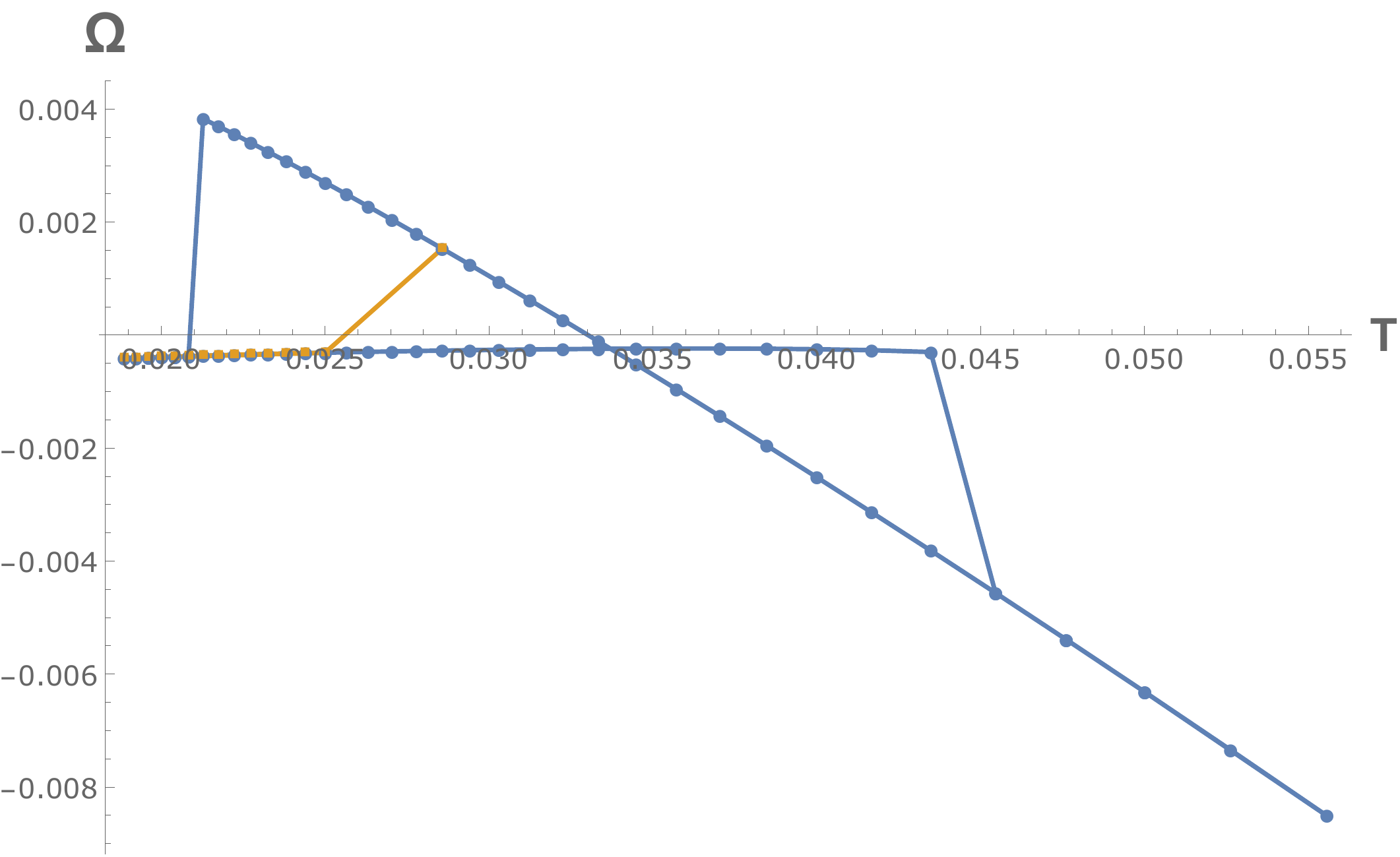}
  \caption{$\mu=0.27$}
  \label{fig:phasetransition_mu027}
\end{subfigure}
\caption{\small{Plot of grand potential $\Omega$ with varying temperature $T$ for different values $\mu$. Phase transition from chaotic phase to integrable phase occurs at different temperatures. (a) The blue curve is for temperature step $\Delta T=0.002$. The orange curve is for $\Delta T=0.001$. The red curve is for inverse temperature steps $\Delta \beta=2$. The transition from integrable phase to chaotic phase occurs at the same temperature $1/\beta\sim1/26$. (b) The blue curve is for $\Delta\beta=2$. The orange curve shows the transition from chaotic to integrable phase when the system is suddenly cooled down from $\beta=35$ to $\beta=40$.}}
\label{fig:phasetransition_mu}
\end{figure}

We observed a chaotic-integrable phase transition. This phase transition has been found previously for the same model and for other similar models \cite{Banerjee_2017,Azeyanagi:2017drg,Maldacena:2018lmt,Choudhury:2017tax}. Our result agrees with the phase diagram in \cite{Azeyanagi:2017drg}. The system is always in the chaotic phase if $\mu < 0.2125$. For $\mu\in]0.2125,0.345[$, there is a finite range of $\beta$ where both the chaotic and integrable phases co-exist. In this temperature range, the system tends to stay in the chaotic if we are cooling down the system slowly from the chaotic phase. While heating up from the integrable phase, the system stays in the integrable phase until a fixed temperature. This results in the hysteresis loop in Figure (\ref{fig:phasetransition_mu024}) and (\ref{fig:phasetransition_mu027}). For $\mu>0.345$, the system is always in a single state and there is no sharp transition.

\begin{figure}[h]
\begin{center}
\includegraphics[width=0.6\textwidth]{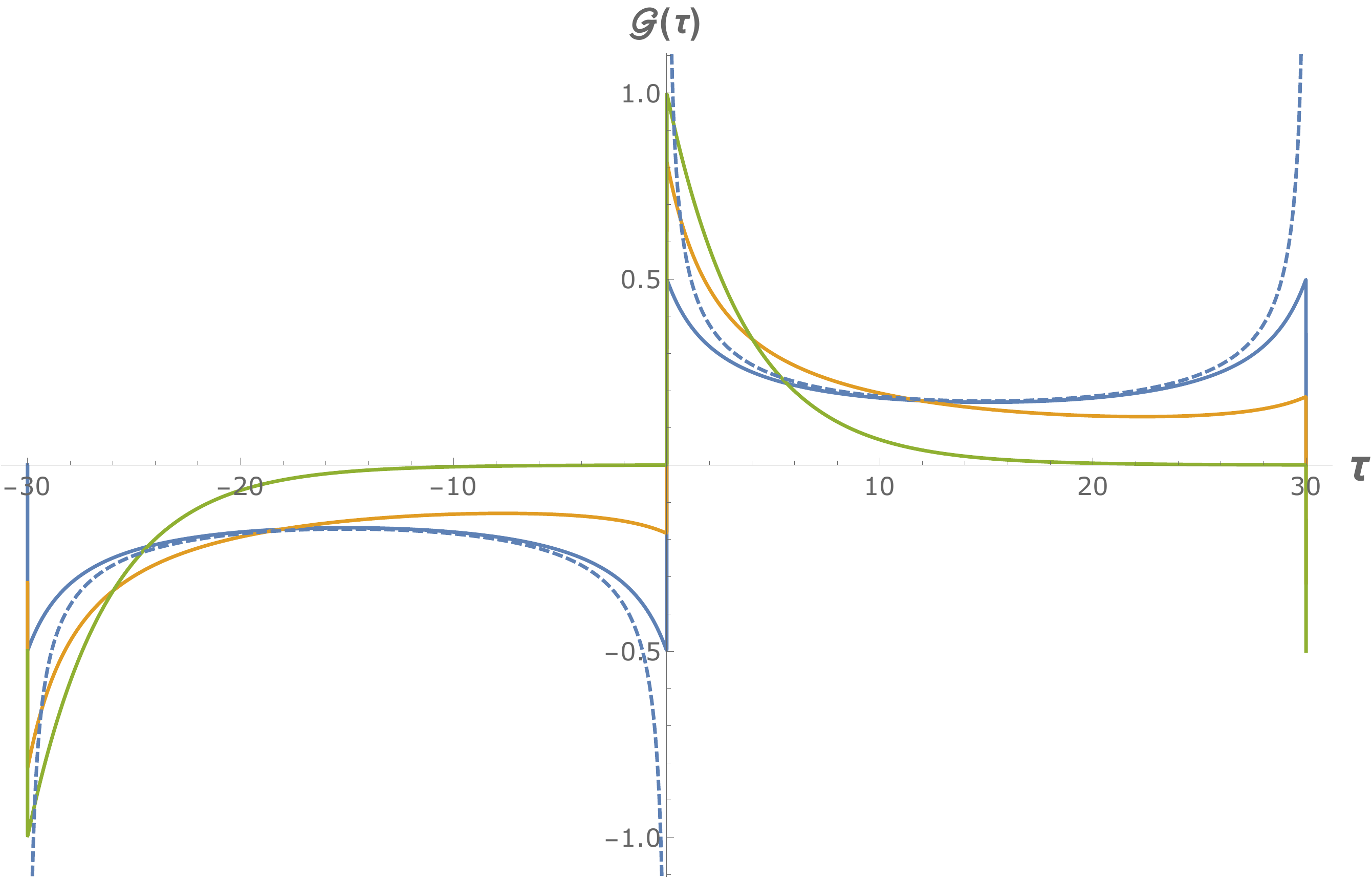}
\caption{\small Thermal propagators at $\beta=30$ in the chaotic phase and integrable phase. The blue curve is for $\mu=0$ in highly chaotic state. The orange curve is for $\mu=0.27$ again in the chaotic state. The green curve is for the same $\mu=0.27$ but in the integrable state. Note that we have chosen the inverse temperature $\beta=30$ to highlight the contrast between the chaotic and the integrable phase. Moreover for $\mu=0.27$, $\beta=30$ is close to the critical point where the grand potentials in both the chaotic and integrable phase are almost equal (see Figure \ref{fig:phasetransition_mu027}).}
\label{fig:therm_prop}
\end{center}
\end{figure}

Even before the explicit calculation of the grand potential, the two different phases can be identified from the propagator $\mathcal{G}(\tau_1,\tau_2)$. In the chaotic phase, the propagator is non-monotonic (say in the range $0<\tau_1-\tau_2<\beta$). While, in the integrable phase, the propagator is monotonic and decays exponentially with time like in the case of free theory (\ref{gtaumu}). This means a gap \cite{Maldacena:2018lmt}. In the chaotic phase with $\mu$ turned on, the Green's function have intermediate behaviour between the two extremes. With $\mu = 0$ and large $\beta$, the mid-section of the imaginary time range is well approximated by
\begin{equation}
\frac{1}{(4\pi J_4^2)^{1/4}}\left(\frac{\pi}{\beta\,\sin[\pi(\tau_1-\tau_2)/\beta]}\right)^{1/2}\
\label{cftlimit}
\end{equation}
which is the conformal limit. Figure \ref{fig:therm_prop} is a plot for the three cases.

As we mentioned above, $\Delta\mathcal{G}$ defined in (\ref{delGim}) is the measure of convergence. $\Delta\mathcal{G}$ decreases rapidly and monotonically if the cooling/heating process keeps the system in the same phase. A signal for an impending phase transition is that $\Delta\mathcal{G}$ will hit a bump and cross it. It would decrease first to some extent and then increase for a while and then decrease and converge rapidly. So, it is important to check if the preset tolerance limit of $\Delta\mathcal{G}$ is small enough to detect the phase transition. But we also would like to note that the phase transition does not depend on the tolerance limit once this quantity is set to a small enough value. We have verified it by running the same cooling and heating process with different tolerance limits.

%

A new feature we observed is that, starting from chaotic phase at a high temperature, a big temperature jump during the cooling process tends to induce the phase transition and take the system to the integrable phase. For example in Figure (\ref{fig:phasetransition_mu027}), a sudden temperature jump from $\beta=35$ to $40$ takes the system to the integrable phase while using small steps of $\delta\beta=1$ keeps the system in the chaotic phase for a longer range of temperature. However this is not observed during the heating process starting from the integrable phase. Starting from the integrable phase for a fixed $\mu$, the transition always happens at same temperature irrespective of the temperature step size used. Note that if one is solely interested in the integrable phase then a closer examination with higher precision is required.

It is worth mentioning here that in the region where both phases coexist the Lyapunov exponent in the integrable phase is non-zero and it gradually increases as we increase the temperature. Whereas outside the hysteresis loop, Lyapunov exponent in the low temperature integrable phase is effectively zero. More details can be found in Section \ref{sec:lyap}.


\section{No phase transition in $(q=2,4)$ model}
\label{sec:q24}
In this section, we will consider the theory with both $q=2$ and $q=4$ interaction without chemical potential.
\begin{eqnarray}
\label{partfunc2_j2}
Z&=&Det\left[\delta(\tau_1-\tau_2)\partial_{\tau_2}+\Sigma(\tau_1,\tau_2)\right]^N\;e^{\left[-\frac{3NJ^2_4}{4}\int_0^{\beta}d\tau_1 d\tau_2 \; \Sigma(\tau_1,\tau_2)\mathcal{G}(\tau_2,\tau_1)\right]}\\
\text{where} && \qquad \Sigma(\tau_1,\tau_2)=-J_2^2\mathcal{G}(\tau_1,\tau_2)+J_4^2\mathcal{G}(\tau_1,\tau_2)^2\mathcal{G}(\tau_2,\tau_1)\nonumber\
\end{eqnarray}
First note that without chemical potential, the charge or occupation number
\begin{equation}
Q=-\lim_{\tau_1\to\tau_2^-}\mathcal{G}(\tau_1,\tau_2)=\frac{1}{2}\
\end{equation}
just as in case of SYK model with Majorana fermions. Moreover, without chemical potential,
\begin{equation}
\mathcal{G}(\tau_1,\tau_2)=-\mathcal{G}(\tau_2,\tau_1)\
\end{equation}
With this relation, the SD equations of SYK model with complex fermions reduce to the SD equations of SYK model with Majorana fermions. The expression of partition function and grand potential are also similar except for the fact that the degrees of freedom is halved for Majorana fermions. So wherever $N$ appears is replaced by $N/2$. Accordingly the results below also applies to SYK model with Majorana fermions.

The $q=2$ interaction is integrable and non-chaotic. So, it would be interesting to check if large value of $J_2$ leads to a sharp transition between chaotic and integrable phases or if it leads to a slow crossover. \cite{PhysRevLett.120.241603} examines the spectrum, spectral form factor and calculates the Lyapunov exponent of this system. It was claimed that there is a phase transition. Using large $q$ limit, it was calculated that the transition would happen at the temperature given by the root of
\begin{equation}
1-\left(\frac{\beta J_2}{\pi}\right)^2\left(\frac{1}{72}+\frac{19-18\log(\pi)}{36\beta J_4}\right)+\mathcal{O}\left(\frac{1}{(\beta J_4)^2}\right)\
\label{j2largeq}
\end{equation}
For $J_4=1$ and $J_2=0.5$, the root is $\beta\approx 55$ or $T\approx 0.018$. For $J_4=1$ and $J_2=2$, the root is $\beta\approx 15$ or $T\approx 0.067$. Below we will work with these values of the coupling strengths.

\begin{figure}[h]
\centering
\begin{subfigure}{.5\textwidth}
  \centering
  \includegraphics[width=.9\linewidth]{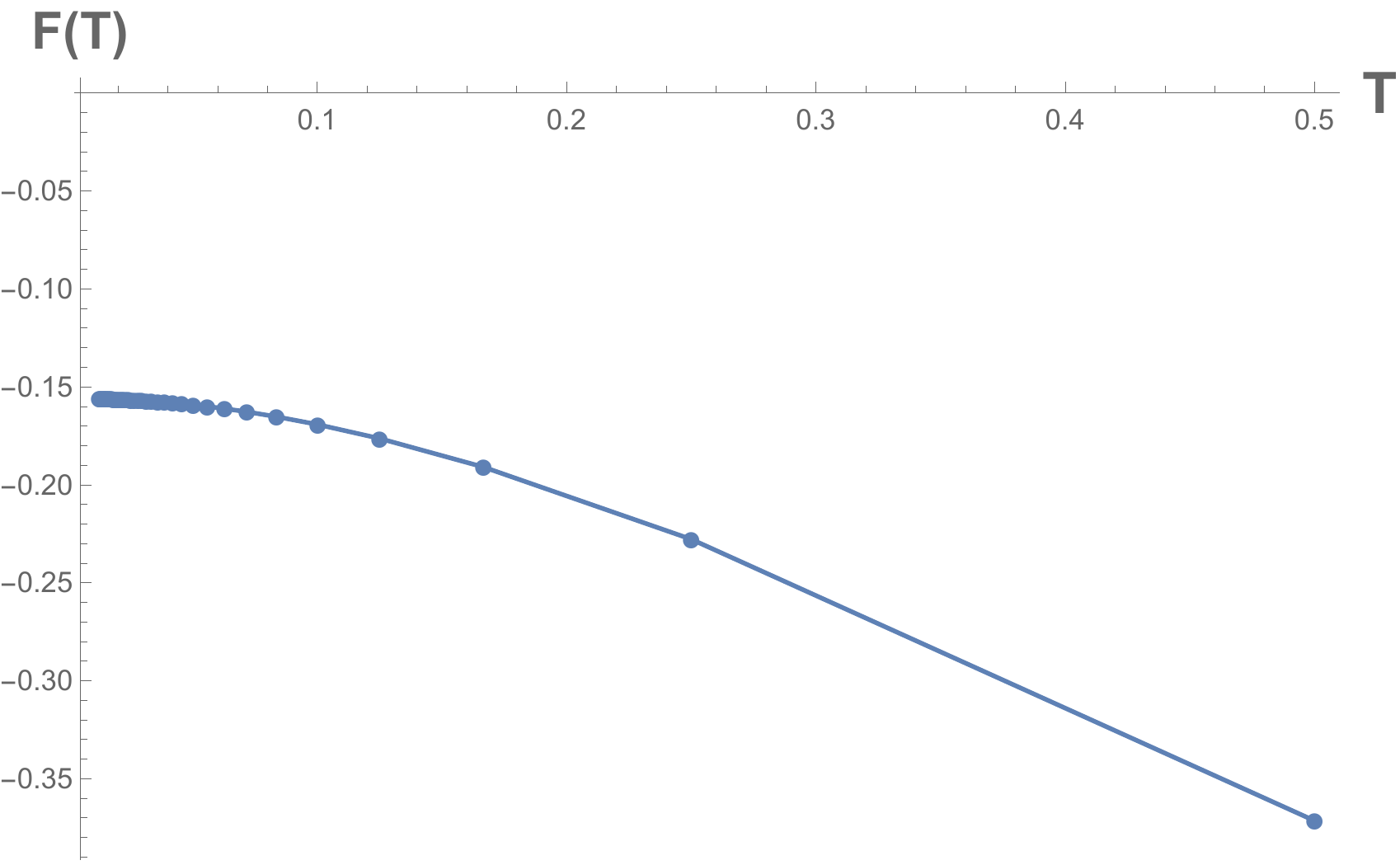}
  \caption{$\mu=0.24$}
  \label{fig:phasetransition_JfJt05}
\end{subfigure}%
\begin{subfigure}{.5\textwidth}
  \centering
  \includegraphics[width=.9\linewidth]{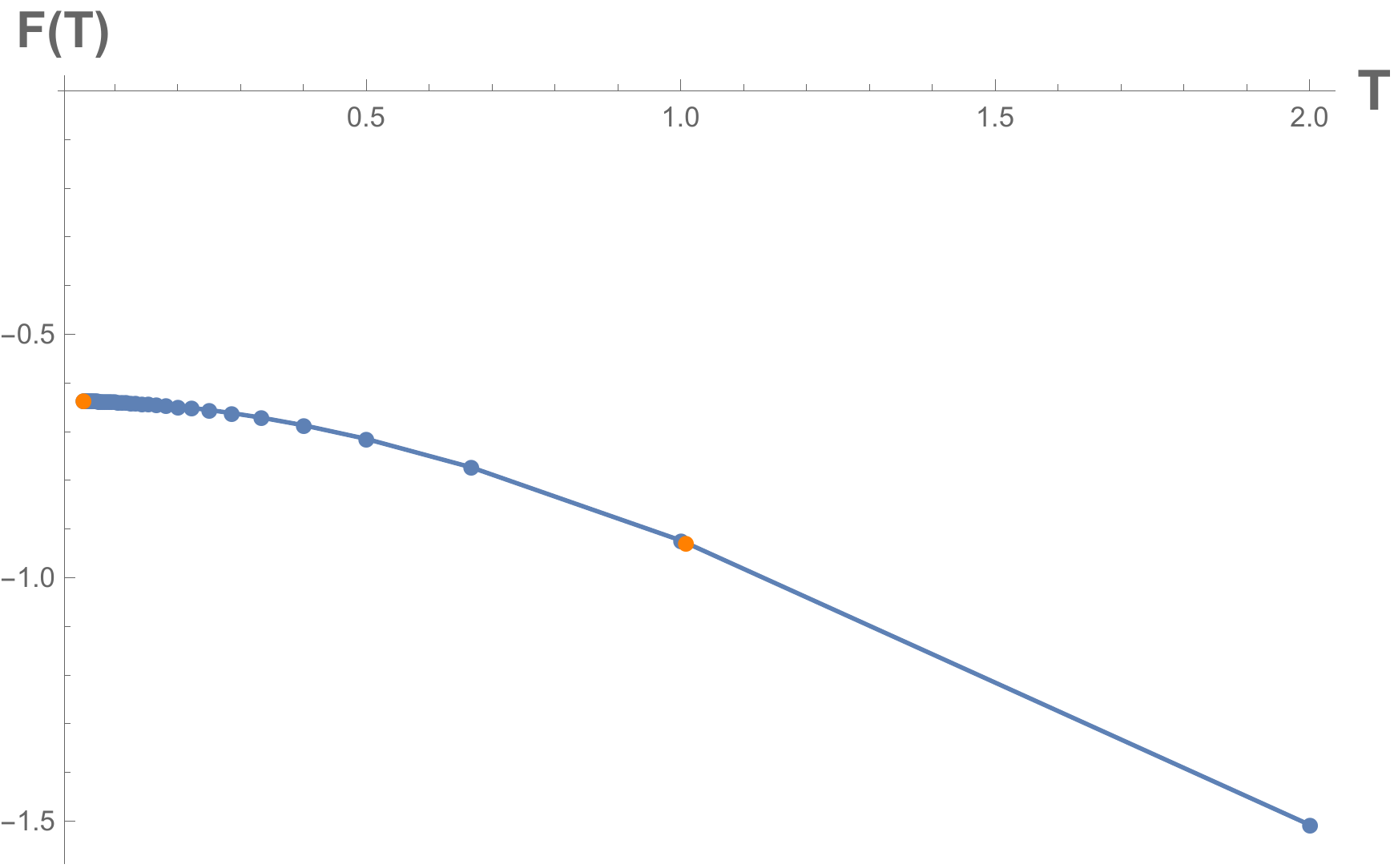}
  \caption{$\mu=0.27$}
  \label{fig:phasetransition_JfJt20}
\end{subfigure}
\caption{\small{Plot of free energy $F$ with varying temperature $T$ for different values $J_2$ and $J_4=1$. No phase transition is observed. (a) $J_2=0.5$ and cooling down in inverse temperature steps of $\Delta\beta=2$ upto $\beta=80$. (b) $J_2=2$ and the blue curve is for inverse temperature steps of $\Delta\beta=0.5$ and cooling down upto $\beta=20$. The orange dots are for cooling down with a big inverse temperature step from $\beta=1$ to $\beta=20$.}}
\label{fig:phasetransition_JfJt}
\end{figure}

Our calculation of the free energy does not show any sharp transition. So, strictly speaking, the system is always in the chaotic phase. Figure \ref{fig:phasetransition_JfJt} are plots of the free energy calculated for different values of $J_2$. We have cooled down the system to temperatures well below the values calculated above. We have also cooled down the system in large temperature steps to see if it leads to a phase transition. Still there is no sharp phase transition. While solving the SD equations, $\Delta\mathcal{G}$ also always decreases rapidly and monotonically.

%

As we have mentioned in section \ref{sec:intro}, this agrees with the result of high precision calculation of the Lyapunov exponent of this system. The Lyapunov exponents are non-zero at temperatures well below the values calculated using (\ref{j2largeq}). This also agrees with the observation that this system always thermalizes.

\section{Complex SYK model in real time formalism}
\label{csykrt}
One can repeat the derivation of the Schwinger-Dyson equations in real time. Or we could use Wick rotation. The real time SD equations are
\begin{gather}
\label{SDrt}
G^R(\omega)=\frac{1}{\omega-\mu-\Sigma^R(\omega)}, \qquad \Sigma^R(t_1,t_2)=\Theta(t_1-t_2)\left(\Sigma^>(t_1,t_2)-\Sigma^<(t_1,t_2)\right)\\
\Sigma^>(t_1,t_2)=G^<(t_2,t_1)G^>(t_1,t_2)^2,\quad \Sigma^<(t_1,t_2)=G^>(t_2,t_1)G^<(t_1,t_2)^2\
\label{sigmart}
\end{gather}
where we have used the conventions in Appendix \ref{conventions}. One could attempt numerical Wick rotation. But we could not do it. The approximating function has to be calculated with extremely high precision which is beyond our computational resource.

Instead, we resort to solving the real time SD equation using an iterative method again. The connecting piece which complete the iterative loop is the Fluctuation-Dissipation relations. They are the expressions of the greater and lesser Green's functions in terms of the spectral function $A(\omega)$.
\begin{eqnarray}
\label{fdt1}
G^>(\omega)&=&-\,\frac{i}{1+e^{-\beta(\eta+\omega)}}\,A(\omega)\\
G^<(\omega)&=&\frac{i}{1+e^{\beta(\eta+\omega)}}\,A(\omega)\
\label{fdt2}
\end{eqnarray}
In Appendix \ref{fdchem}, we have derived these relations. The steps involved in solving the real time SD equations are as follows.

\begin{enumerate}
\item Just like the case of imaginary time formalism, the initial values of the iterations is the spectral function of the solvable $q=2$ SYK model. The real time solution of $q=2$ SYK model is
\begin{equation}
A(\omega)=\frac{1}{J_2}\,\sqrt{4J^2_2-\omega^2}, \qquad \omega \in \{-2J^2_2,2J_2^2\}\
\end{equation}
\item $G^>(\omega)$ and $G^<(\omega)$ are calculated using the Fluctuation-Dissipation relations. Fourier transforms give $G^>(t_1,t_2)$ and $G^<(t_1,t_2)$.
\item The next step is to calculate $\Sigma^R(\omega)$. One could directly use convolutions to calculate $\Sigma^R(\omega)$ from $G^>(\omega)$ and $G^<(\omega)$. But the three integrals during convolution are computationally more expensive. So it is better to use the expression of real time $\Sigma(t_1,t_2)$ and perform a Fourier transform.
\item Next we calculate $G^R(\omega)$ using equation (\ref{SDrt}) and the values of $\Sigma^R(\omega)$ calculated above.
\item Using the calculated value of $G^R(\omega)$, we calculate the new $A(\omega)$. The relation between $G^R(\omega)$ and $A(\omega)$ defined in (\ref{defA}) is
\begin{equation}
A(\omega)=-2\;\text{Im}\,G^R(\omega)\
\end{equation}
\item Finally we can calculate $G^>(\omega)$ and $G^<(\omega)$ using the new $A(\omega)$. After this, we go back to step (2) and repeat the iteration until the desired tolerance limit is reached. The convergence is checked by calculating the difference of the spectral function of the new iteration and the previous iteration.
\begin{equation}
\Delta A=\sum_{\omega}|A(\omega)-A_{prev}(\omega)|\
\end{equation}
\end{enumerate}
We also find the phase transition in the real time formalism. We also observed that the phase transition happens at roughly the same temperature as the one observed in imaginary time calculation for the same $\mu$. Moreover, $\Delta A$ crosses a bump to go from one phase to the other just like $\Delta\mathcal{G}$ in the case of imaginary time formalism.

With $\mu$ or $\eta$ turned on, the charge or occupation number $Q=-i\lim_{t_1\to t_2} G^<(t_1,t_2)$ is no longer $\frac{1}{2}$. As we expect, the occupation number is same for the same numerical value of $\mu$ (in system with Hamiltonian $\tilde{H}_{SYK}$) and $\eta$ (in system with  Hamiltonian $H_{SYK}$). But at finite temperature in which we are working, analytic relation between $Q$ and $\mu$ or $\eta$ is so far lacking.

We solve the real time SD equations by discretizing time into 10000 intervals of size $dt = 0.05$. We took the frequency range from $-3000$ to $3000$ with interval size $d\omega=0.001$.

\begin{figure*}[h]
        \centering
        \begin{subfigure}[b]{0.475\textwidth}
            \centering
            \includegraphics[width=\textwidth]{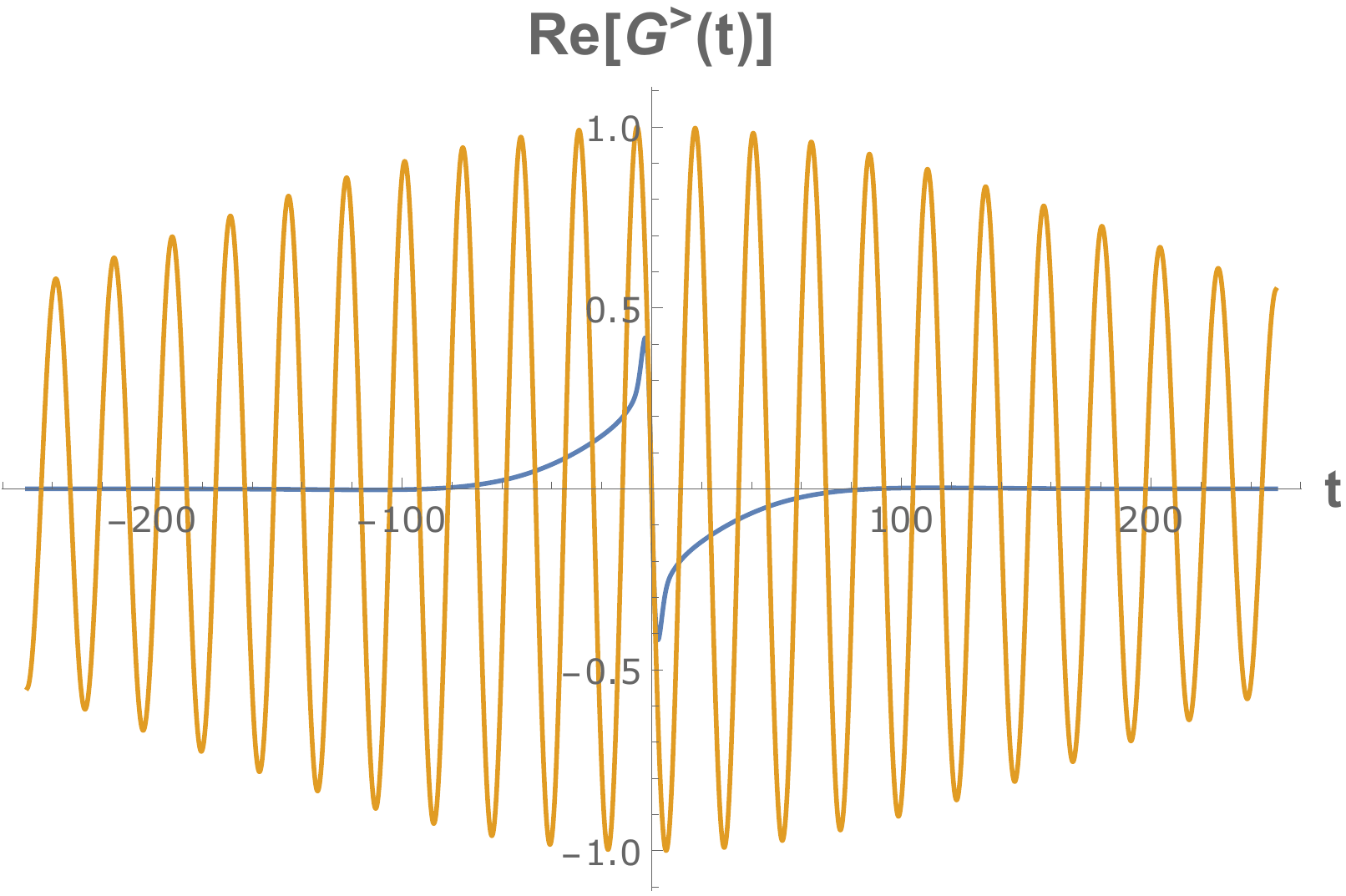}
            \caption[]%
            {{\small $\mu=0.27$. Real parts of $G^>(t)$.}}    
            \label{fig:re_mu}
        \end{subfigure}
        \hfill
        \begin{subfigure}[b]{0.475\textwidth}  
            \centering 
            \includegraphics[width=\textwidth]{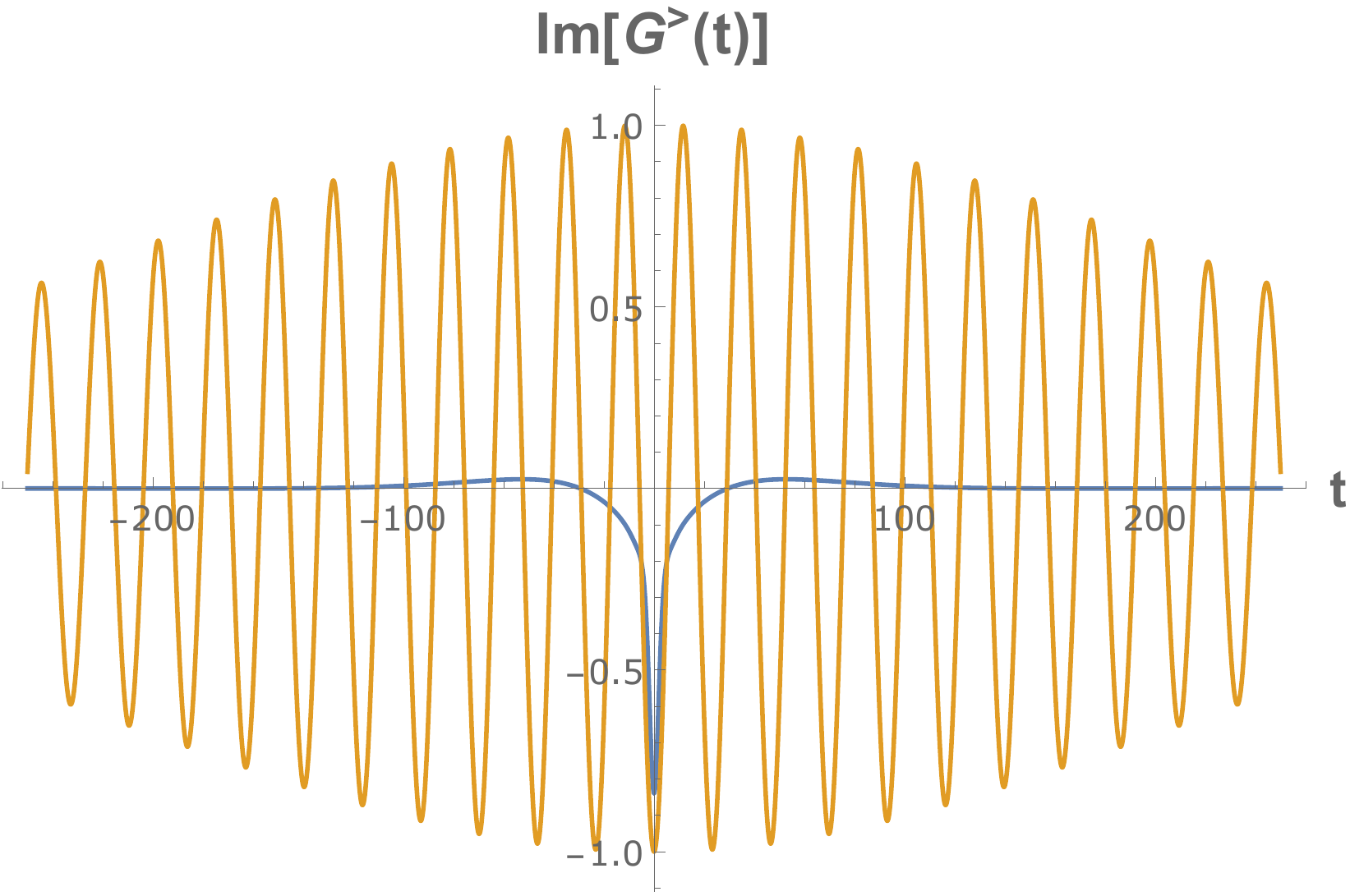}
            \caption[]%
            {{\small $\mu=0.27$. Imaginary parts of $G^>(t)$.}}    
            \label{fig:im_mu}
        \end{subfigure}
        \vskip\baselineskip
        \begin{subfigure}[b]{0.475\textwidth}   
            \centering 
            \includegraphics[width=\textwidth]{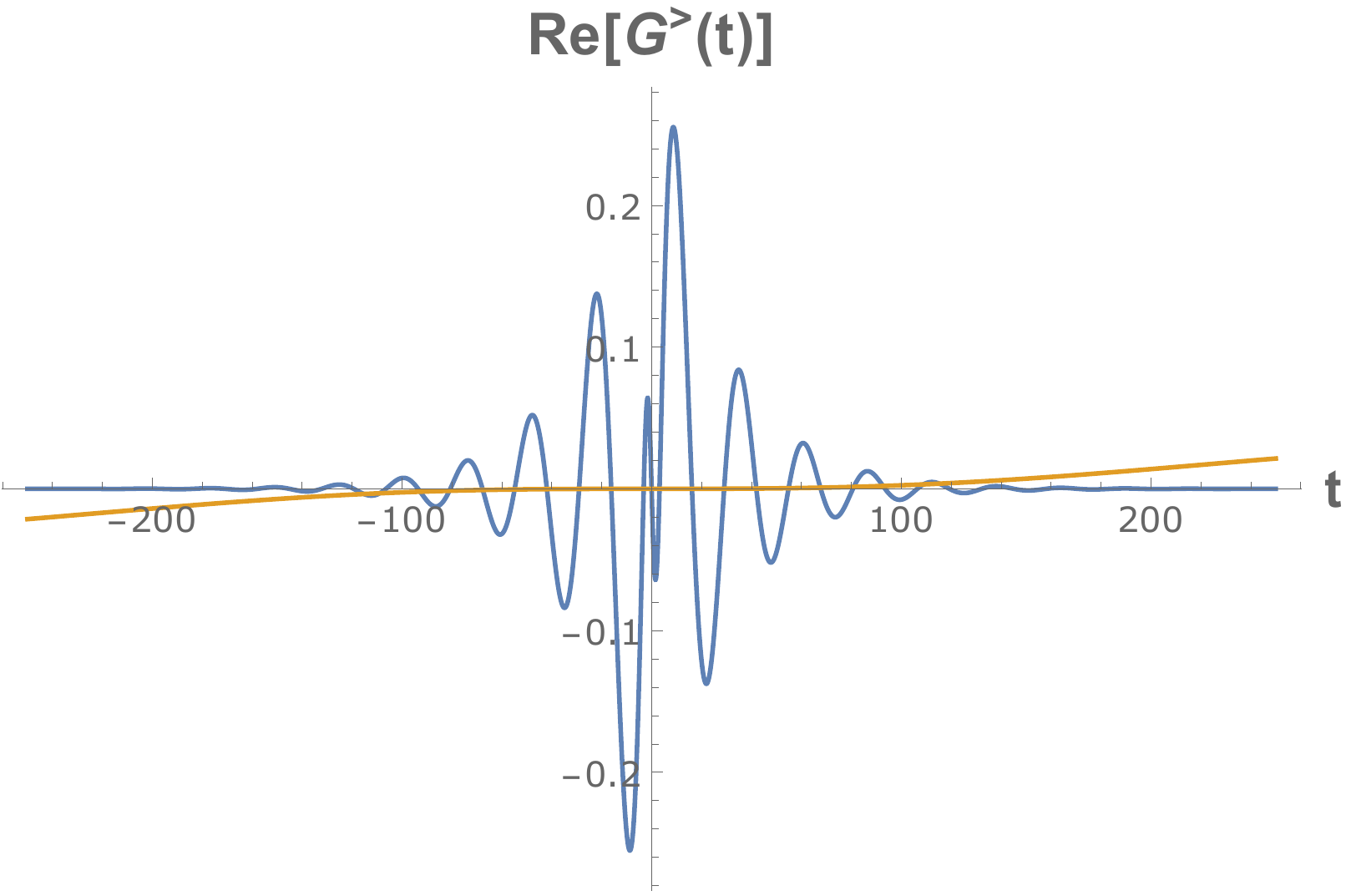}
            \caption[]%
            {{\small $\eta=0.27$. Real parts of $G^>(t)$.}}    
            \label{fig:re_eta}
        \end{subfigure}
        \quad
        \begin{subfigure}[b]{0.475\textwidth}   
            \centering 
            \includegraphics[width=\textwidth]{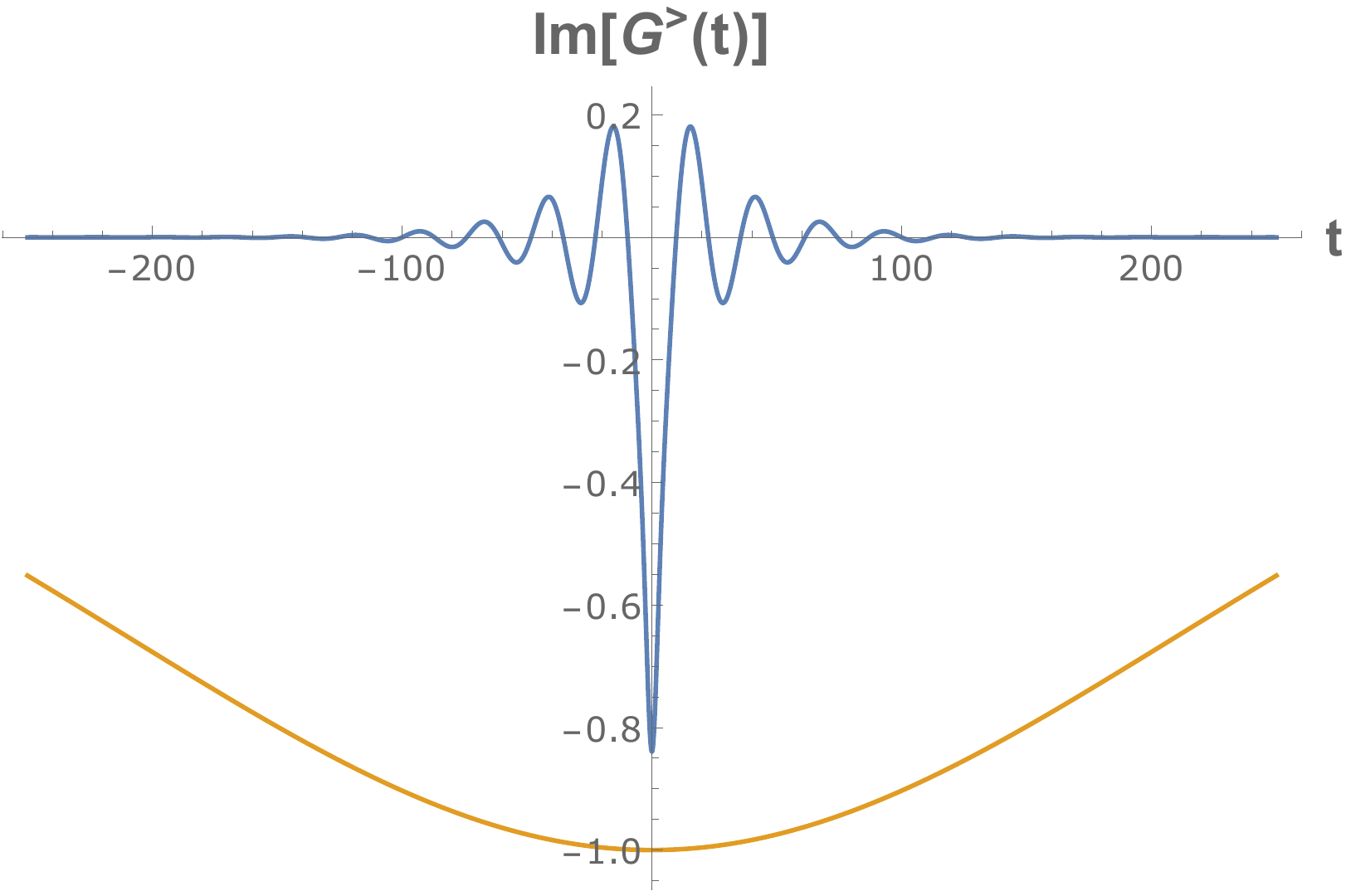}
            \caption[]%
            {{\small $\eta=0.27$. Imaginary parts of $G^>(t)$.}}    
            \label{fig:im_eta}
        \end{subfigure}
        \caption[]{\small{Greater Green's functions $G^>(t=t_1-t_2)$ for $\beta=40$ in the two different phases. The blue curves are the values in the chaotic state while the orange curves are the values in the integrable phase. (a) is the real parts with $\mu=0.27$, (b) is the imaginary parts with $\mu=0.27$, (c) is the real parts with $\eta=0.27$ and (d) is the imaginary parts with $\eta=0.27$. Note that the plots of $G^>(t)$ in the integrable phase are for illustrative purpose. They have significant finite-$t$ effect.}} 
        \label{fig:gg_mueta}
\end{figure*}

The lesser (greater) Green's function $G^{<(>)}(t)$ is very small for positive (negative) mass $\mu$ or chemical potential $\eta$ in both the chaotic phase and integrable phase. In free theory, they are exponentially suppressed. So taking $\mu,\eta>0$, we have plotted only the greater Green's functions in Figure \ref{fig:gg_mueta}. The plots of $G^>(t)$ in the integrable phase are close to the value of $G^>(t)$ for free theories given in (\ref{Ggreater1},\ref{Ggreater2}). This again confirms that in the integrable phase the theory is a weakly interacting theory. Also note that the relation between the $G^>(t)$ with the mass $\mu$ and with chemical potential $\eta=\mu$ given in (\ref{Grelation}) can be verified from the plots.

%
%

We also find that cooling down with big temperature steps also induce the transition from the chaotic phase to the integrable phase. But for this bigger temperature jumps are required compare to the imaginary time calculation. For example in Figure \ref{fig:phasetransition_mu027} for $\mu=0.27$, a temperature jump from $\beta=35$ to $40$ takes the system from the chaotic phase to the integrable. But in present case of the real time calculation, $\beta=1$ to $40$ is required for the same transition with $\eta=0.27$.
\section{Calculation of Lyapunov exponent}
\label{sec:lyap}
\begin{figure}[H]
\begin{center}
\includegraphics[width=0.8\textwidth]{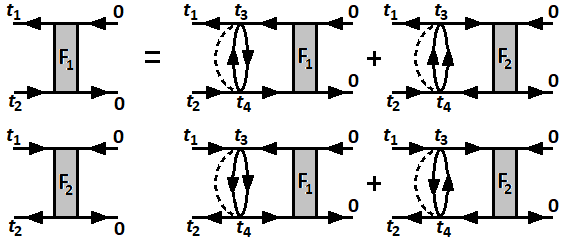}
\caption{\small Feynman diagrams for OTOC calculation.}
\label{fig:feynman}
\end{center}
\end{figure}

The Lyapunov exponent is calculated by diagonalizing the retarded kernel \cite{Banerjee_2017}. The recursive Feynman diagram for OTOC is shown is Figure (\ref{fig:feynman}). The upper legs are $i\beta/2$ imaginary time away from the lower legs.
The eigenvalue problem is,
\begin{gather*}
\begin{bmatrix}
F_1\\
F_2
\end{bmatrix}=k\begin{bmatrix}
K_{11} & K_{12}\\
K_{21} & K_{22}
\end{bmatrix}
\begin{bmatrix}
F_1\\
F_2
\end{bmatrix}\\
K_{11}=2J_4^2G^A(t_{31})G^R(t_{24})G^+(t_{43})G^-(t_{34})\\
K_{12}=-J_4^2G^A(t_{31})G^R(t_{24})G^+(t_{43})G^+(t_{43})\\
K_{21}=-J_4^2G^R(t_{13})G^A(t_{42})G^-(t_{34})G^-(t_{34})\\
K_{22}=2J_4^2G^R(t_{13})G^A(t_{42})G^-(t_{34})G^+(t_{43})\
\end{gather*}
where $G^+(t)=G^>(t-i\beta/2)$ and $G^-(t)=G^<(t+i\beta/2)$ ($=-G^+(t)$).
We solve this equation in frequency space and look for eigenvalue $k=1$. This is done by tuning $\lambda$ in the OTOC ansatz
\begin{equation}
F_{1,2}(t_1,t_2)=e^{\lambda (t_1+t_2)/2}f_{1,2}(t_1,t_2)\
\end{equation}
After some algebraic steps the final equation is
\begin{gather}
\begin{bmatrix}
f_1(\omega) \\
f_2(\omega)
\end{bmatrix}=k\begin{bmatrix}
K_{11}(\omega,\omega') & K_{12}(\omega,\omega') \\
K_{21}(\omega,\omega') & K_{22}(\omega,\omega')
\end{bmatrix}
\begin{bmatrix}
f_1(\omega') \\
f_2(\omega')
\end{bmatrix}\\
K_{11}(\omega,\omega')=\frac{2J_4^2}{2\pi}\,G^A\left(-\omega-\frac{i\lambda}{2}\right)G^R\left(-\omega+\frac{i\lambda}{2}\right)G^{+-}(\omega'-\omega)\\
K_{12}(\omega,\omega')=-\frac{J_4^2}{2\pi}\,G^A\left(-\omega-\frac{i\lambda}{2}\right)G^R\left(-\omega+\frac{i\lambda}{2}\right)G^{++}(\omega'-\omega)\\
K_{21}(\omega,\omega')=-\frac{J_4^2}{2\pi}\,G^R\left(\omega+\frac{i\lambda}{2}\right)G^A\left(\omega-\frac{i\lambda}{2}\right)G^{--}(\omega'-\omega)\\
K_{22}(\omega,\omega')=\frac{2J_4^2}{2\pi}\,G^R\left(\omega+\frac{i\lambda}{2}\right)G^A\left(\omega-\frac{i\lambda}{2}\right)G^{-+}(\omega'-\omega)\
\end{gather}
$G^{+-}(\omega), G^{++}(\omega), G^{--}(\omega)$ and $G^{-+}(\omega)$ are the Fourier transforms of $G^+(t)G^-(-t)$, $G^+(t)^2, G^{-}(t)^2$ and $G^-(t)G^+(-t)$ respectively.
We solve this equation numerically using BLAS and LAPACK libraries in FORTRAN.
As mentioned in the previous section, we took the frequency range from $-3000$ to $3000$ with interval size $d\omega=0.001$. The diagonalization of the $12002 \times 12002$ real matrix takes of the order of 10 seconds in a modern desktop computer.

The largest eigenvalue $k$ is the one of interest for us. We have to set it to 1 by tuning $\lambda=\lambda_L$, the Lyapunov exponent for the system of interest. We find that $\lambda$ and the largest eigenvalue (all other eigenvalues also) has an inverse relation. Increasing $\lambda$ decreases the eigenvalues and vice versa. So, we do not have to search a large range of $\lambda$. We used bisection method after finding two values of $\lambda$ for which $(k-1)$ has opposite signs.

\begin{figure}[h]
\begin{center}
\includegraphics[width=0.6\textwidth]{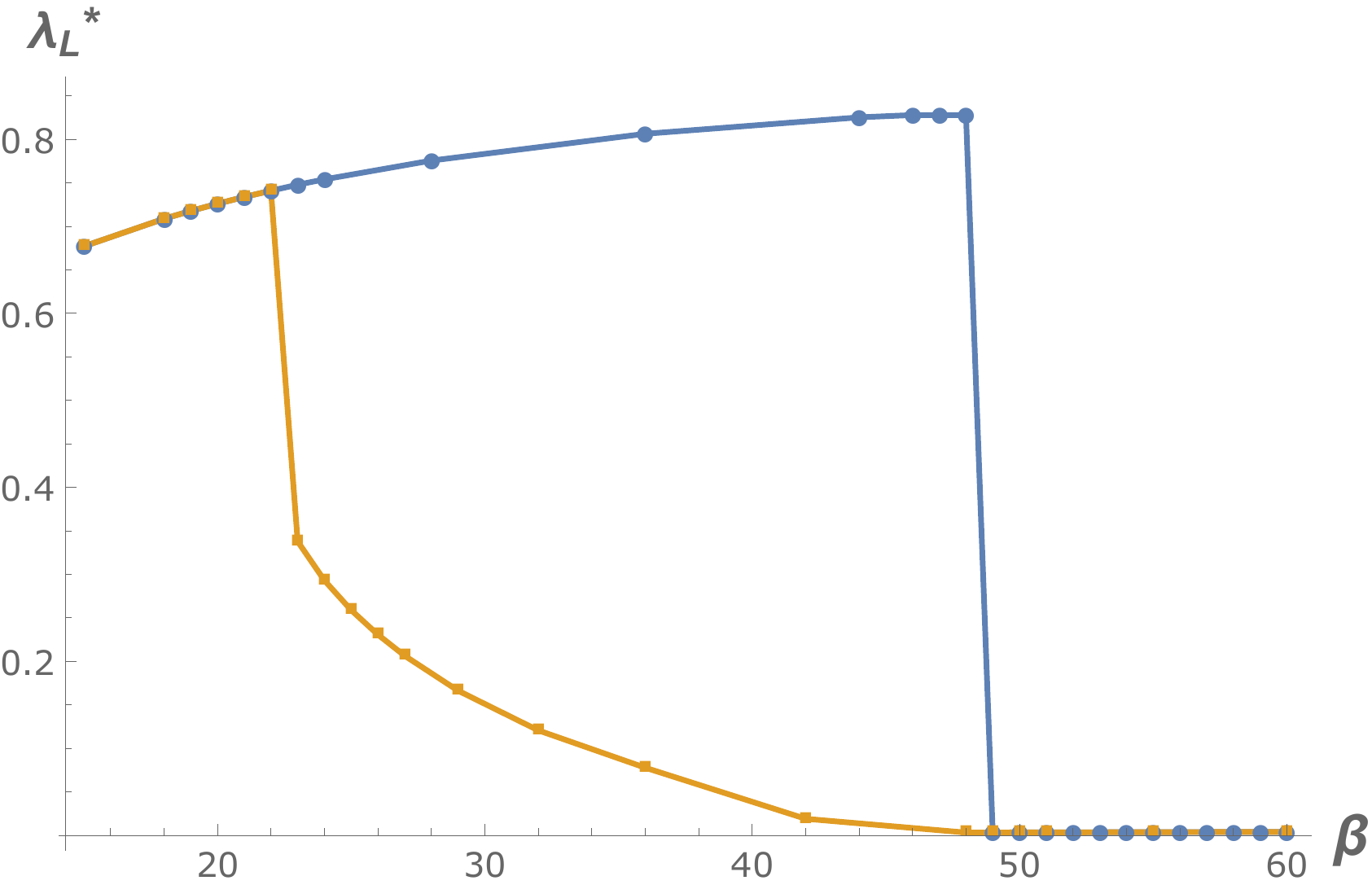}
\caption{\small Normalized Lyapunov exponent in the two different phases for $\eta=0.27$. The blue curve corresponds to the cooling process starting from the chaotic phase. The orange curve corresponds to the heating process starting from the integrable phase. Note that the Lyapunov exponent of the integrable phase is non-zero inside the hysteresis loop.}
\label{fig:phaselyap}
\end{center}
\end{figure}

We calculate the Lyapunov exponent for the systems with Hamiltonians $H_{SYK}$ and $\tilde{H}_{SYK}$ in different states with or without chemical potentials turned on. The results are as follows
\begin{enumerate}
\item The phase transition is also manifested in the Lyapunov exponent as shown in Figure \ref{fig:phaselyap}. $\lambda_L^*$ has a large value in the chaotic phase and it increases with decreasing temperature. At the transition point, it sharply goes to a very small value. Even in the integrable phase, it is non-zero for relatively high temperature especially inside the hysteresis loop where both the two phases can exist.
\item We find that the chemical potential suppresses the Lyapunov exponent exponentially.
Figure (\ref{fig:b100b200}) is the plot of the Lyapunov exponent as a function of the chemical potential for fixed values of $\beta$. It fits an exponential function very well for small values of the chemical potential.
\item We find that the upper bound of the Lyapunov exponent with chemical potential is still $2\pi/\beta$. This can be seen from Figure (\ref{fig:lyap_plots_beta}) where we plot the Lyapunov exponents for different values of $\mu$ including $\mu=0$. For a clearer picture, Figure (\ref{fig:lyap_ratio_plots}) is the plot of the ratio of $\lambda^*_L$ for non-zero $\mu$ and for $\mu=0$. The ratio converses to $1$.
\begin{figure}[h]
\centering
\begin{subfigure}{.5\textwidth}
  \centering
  \includegraphics[width=.9\linewidth]{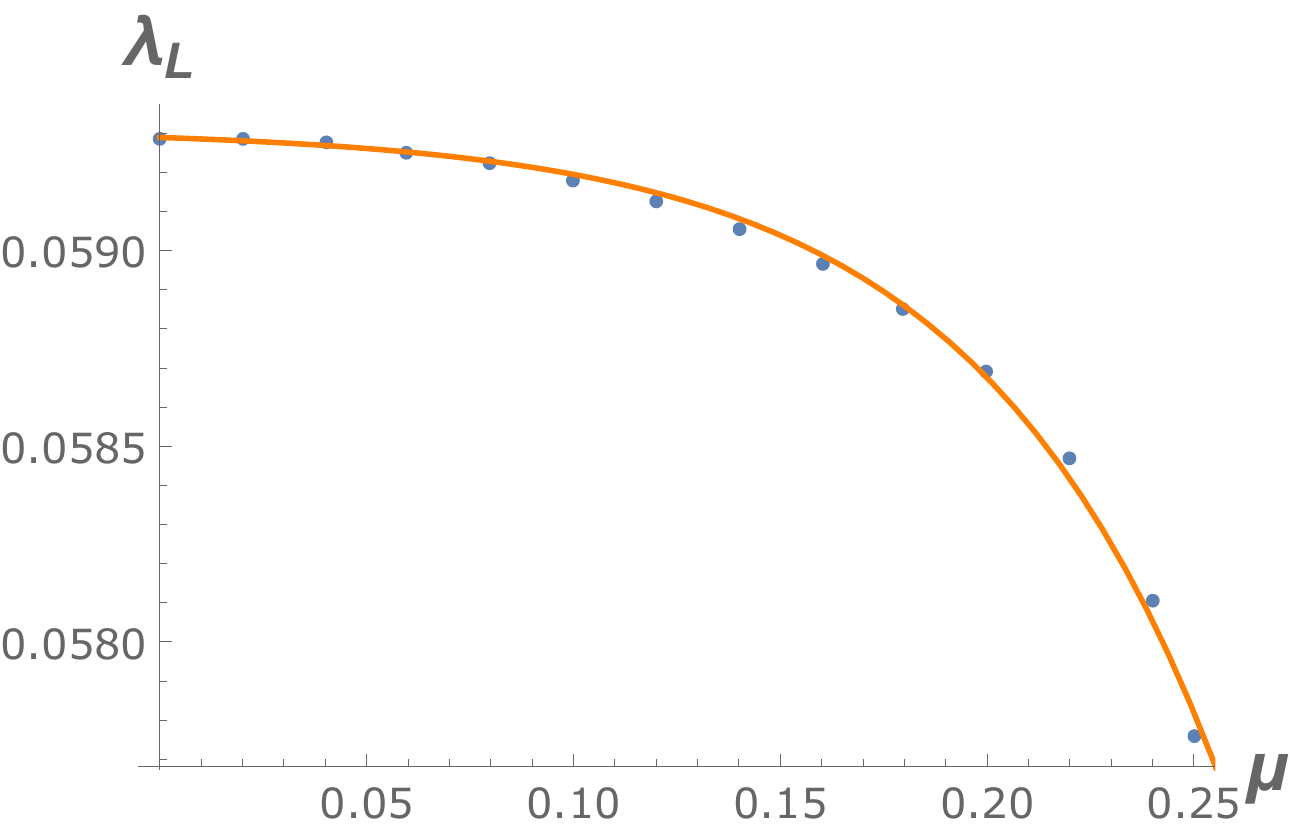}
  \caption{$\beta=100$}
  \label{fig:b100}
\end{subfigure}%
\begin{subfigure}{.5\textwidth}
  \centering
  \includegraphics[width=.9\linewidth]{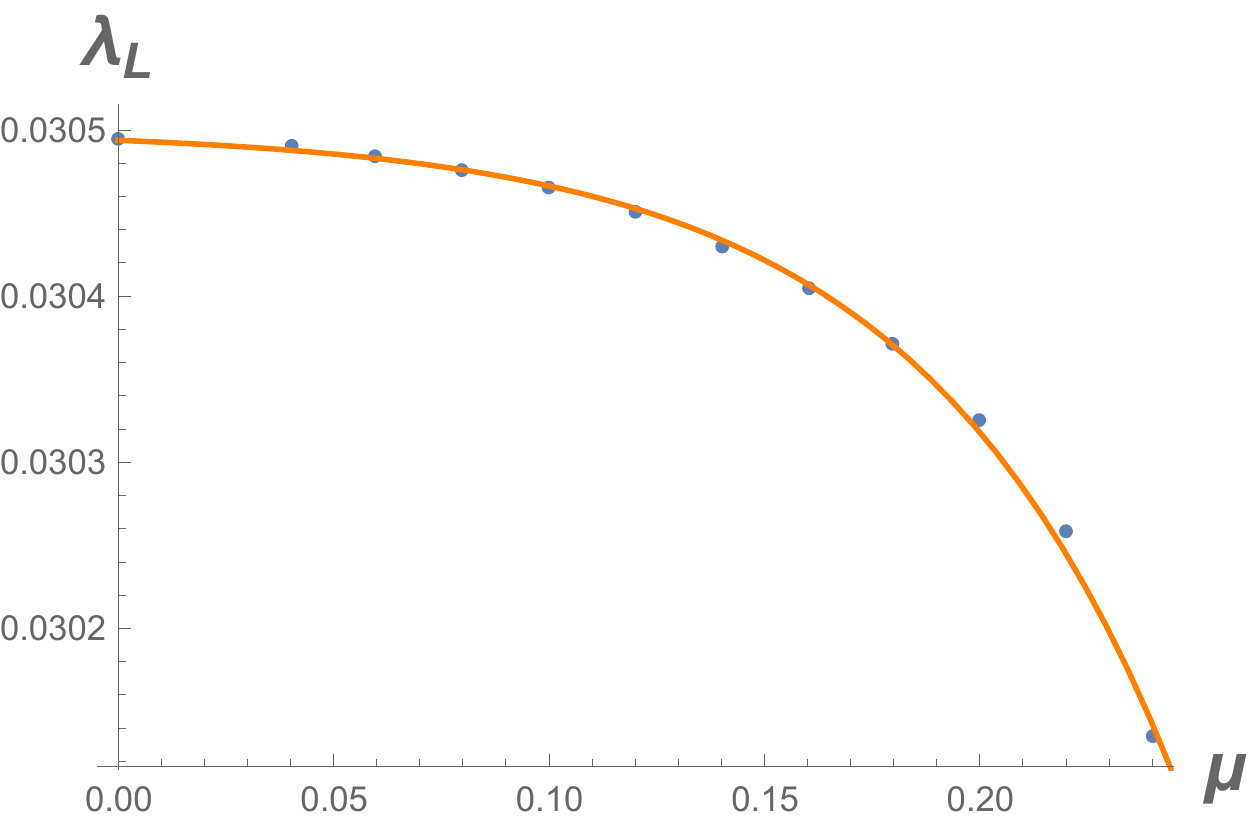}
  \caption{$\beta=200$}
  \label{fig:b200}
\end{subfigure}
\caption{\small{Lyapunov exponent as a function of the chemical potential for fixed temperature. The plots fit exponential functions very well. The fits for (a) is $0.05931\, -0.0000209 \times e^{17.06\;\mu}$ and (b) is $0.03050\, -6.245\times 10^{-6}\; e^{16.86 \;\mu}$.}}
\label{fig:b100b200}
\end{figure}

\begin{figure}[h]
\centering
\begin{subfigure}{.5\textwidth}
  \centering
  \includegraphics[width=.9\linewidth]{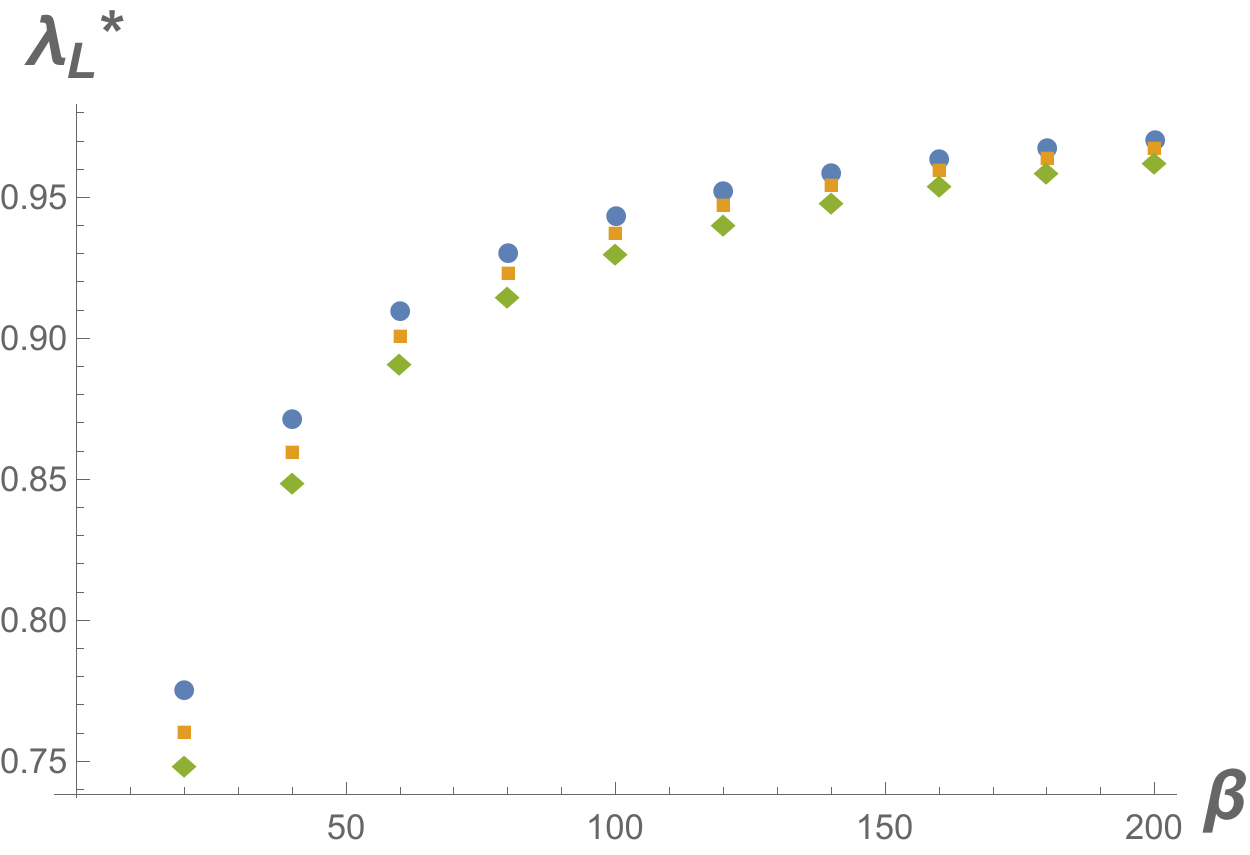}
  \caption{}
  \label{fig:lyap_plots_beta}
\end{subfigure}%
\begin{subfigure}{.5\textwidth}
  \centering
  \includegraphics[width=.9\linewidth]{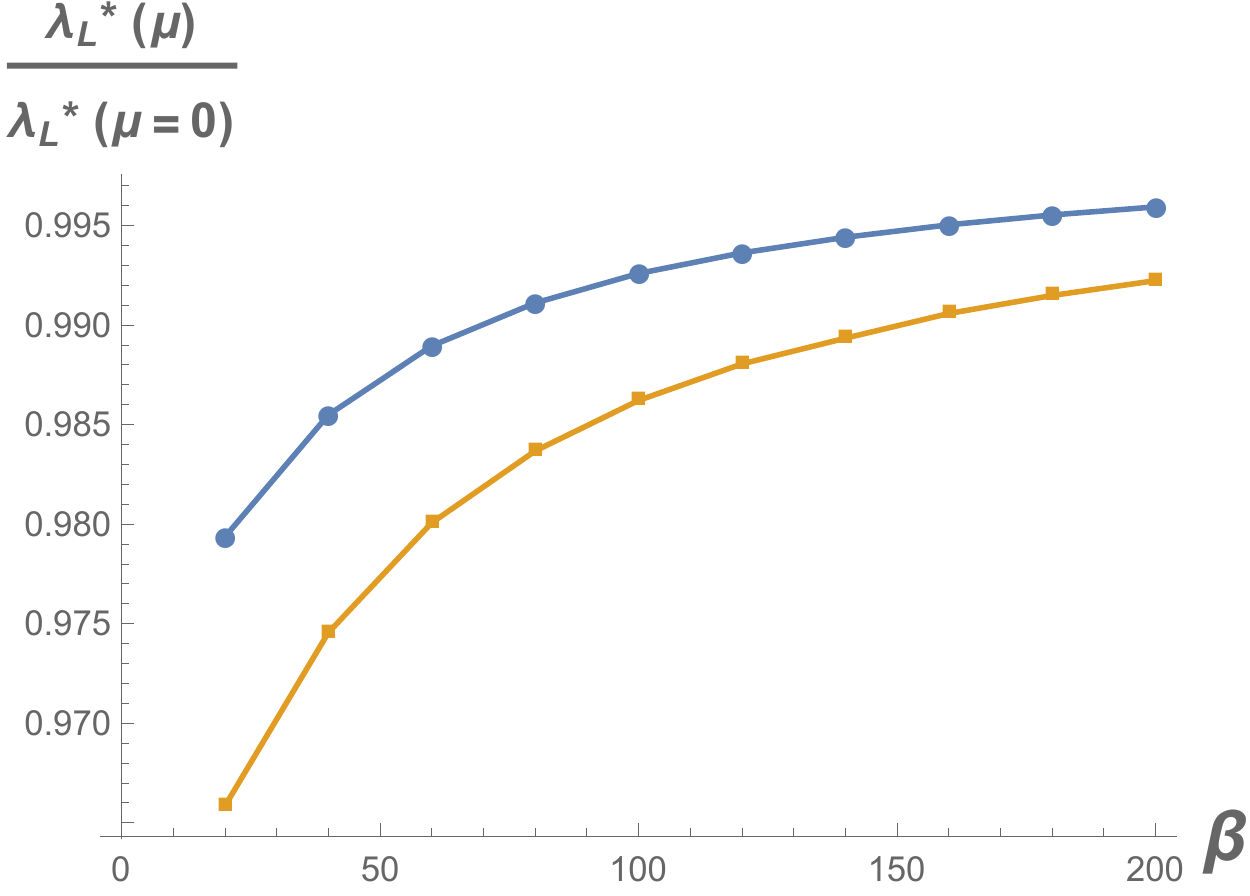}
  \caption{}
  \label{fig:lyap_ratio_plots}
\end{subfigure}
\caption{\small{(a) The Lyapunov exponent for different fixed $\mu=0,0.18,0.22$ as a function of the inverse temperature $\beta$. (b) The ratio of the Lyapunov exponent for non-zero $\mu$ and for $\mu=0$. The ratio converges to 1. This shows that the upper bound of the Lyapunov exponent is still $2\pi/\beta$ even with non-zero chemical potential.}}
\label{fig:lyap_ratio}
\end{figure}
\item In the purely integrable phase outside the hysteresis loop, we find that the normalized Lyapunov exponent $\lambda_L^*$ is below $10^{-6}$ for $\mu=0.28$ and $\beta=90.9$.
\end{enumerate}

\subsection{Comparison with large q result}
\begin{figure}[h]
\begin{center}
\includegraphics[width=0.6\textwidth]{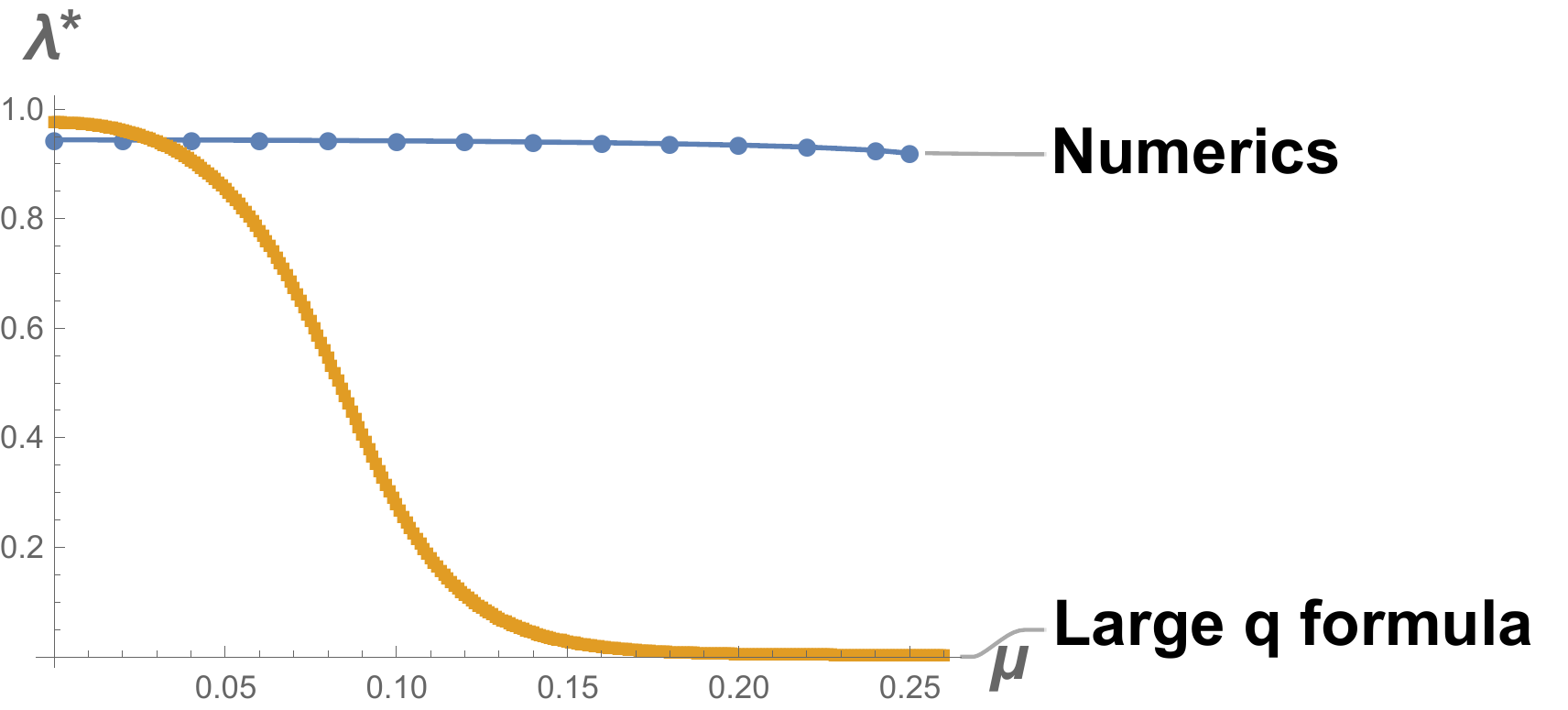}
\caption{\small{Numerical results for the Lyapunov exponent for $q=4$ interaction as against analytical results for large $q$.}}
\label{fig:largeq}
\end{center}
\end{figure}

Here we will compare our results with the large q result of \cite{Bhattacharya:2017vaz}. The scaled Lyapunov exponent $\lambda_L^*$ is given by
\begin{equation}
\pi \lambda_L^* = \beta\tilde{J}_q\,\cos(\pi\lambda_L^*/2), \qquad \text{where} \qquad \tilde{J}_q=\sqrt{\frac{qJ_q^2}{2(2+2\cosh(\mu\beta))^{q/2-1}}}\
\label{largeq}
\end{equation}
For $\beta\tilde{J}<1$, the solution is a relative of the Dottie number \cite{dottienum}. Dottie number is the solution of $x=\cos(x)$. It is a universal attracting fixed point in the real line which one can simply check by repeatedly taking $\cos$ function of any given real number. Similarly, for the above equation of $\lambda_L^*$, the solutions are calculated by repeatedly taking $\beta\tilde{J}_4\,\cos(\pi\lambda_L^*/2)$. The solutions are transcendental numbers. With $\beta=100, J_4=1$, $\beta\tilde{J}_4 \to 1^-$ for $\mu\sim 0.0990339$. For smaller $\mu$ or larger $\beta$, $\beta\tilde{J}_4$ is greater than 1 and the equation cannot be solved by the above iterative method. The solutions are searched using bisection method. There is only one unique solution for a given set of parameters $J_4, \beta, \mu$ and $q$. Figure (\ref{fig:largeq}) is the plot comparing our numerical result with the large q formula (but putting $q=4$ in the formula). We find that the Lyapunov exponent calculated using the formula (\ref{largeq}) are much more highly suppressed.

\section{Conclusions}
\label{cnd}
In this work, we study the chaotic and integrable phases of SYK model with chemical potential. In imaginary time formalism we explicitly calculate the partition function using a novel numerical technique. The chaotic phase is characterized by a linearly increasing grand potential with decreasing temperature. The integrable phase is characterized by a constant grand potential as a function of the temperature. The system undergoes a sharp transition between the two phases.

We have also shown that $q=2$ SYK term does not lead to a sharp phase transition. This result agrees with the observation in \cite{Bhattacharya:2018fkq} that even in the presence of very strong $q=2$ coupling the system always thermalizes. This was shown by performing a quantum quench in a system with the Hamiltonian given by (\ref{Hj2}). One uses a time-dependent $q=2$ or $q=2$ or $q=6$ coupling to perturb the system. This takes the system out of equilibrium. One solves numerically the equations of motion of the Green's functions called Kadanoff-Baym equations. Still it would be interesting to see if the non-thermalizing path is the most favourable path by calculating the transition amplitude. It has also been shown that the Lyapunov exponent is never effectively zero even when the $q=2$ coupling is very large. In contrast, we have found that in the purely integrable phase due to chemical potential, the Lyapunov exponent is effectively zero.

In real time formalism, we calculate the Lyapunov exponent $\lambda_L^*$ of the system in the two different phases. $\lambda_L^*$ is large in the chaotic phase. At the transition point, it goes to a very small value in the integrable phase. But this does not means that $\lambda_L^*$ is zero in the integrable phase. It can be relatively large at higher temperature. However, $\lambda_L^*$ is effectively zero in the purely integrable phase at low temperatures.

These results point towards the idea that thermalization can be state-dependent. 
It is expected that a system in the chaotic state would thermalize after a perturbation. But a system in the integrable state is not expected to thermalize unless the perturbation is strong enough to take the system to the chaotic state. 

Rigorous numerical calculations in simple models are available on the gravity side. Consider the hard wall model of \cite{Craps:2014eba}. The gravity theory has a hard wall in the bulk. Unless a large enough energy is pump in to create a big black hole with which the horizon swallows up the hard wall, there is no black hole formation.

Our numerical calculation of the OTOC has shown that the chemical potential suppresses the Lyapunov exponent exponentially when we consider the system at finite temperature. In the low temperature limit, the Lyapunov exponent still tends towards the upper bound of (\ref{chaosbound}). We have also shown that a mass term in the Hamiltonian of SYK model does not suppress the Lyapunov exponent.

As future direction of this work, it would to interesting to verify that the integrable state does not thermalize. The chaotic state is expected to thermalize but it would be interesting to see how the chemical potential affects the thermalization process. We expect that the chemical potential would slow down the thermalization process. Another direction is to calculate the operator growth rate with chemical potential. Operator growth rate of SYK model without chemical potential has been calculated in \cite{Roberts:2018mnp, Qi:2018bje}. 

\section*{Acknowledgement}
The author thanks Dileep Jatkar, Ritabrata Bhattacharya and Tousik Samui for helpful discussions. The author also thanks Sayantani Bhattacharyya, Yogesh Srivastava and Tarun Sharma for helpful commments.

\appendix

\section{Conventions}
\label{conventions}
Our definition of the Green's functions are given below. $\mathcal{G}$ is the thermal propagator. $G$'s are real time Green's functions. $\psi^+$ means the operator is inserted in the upper segment of the Keldysh contour while $\psi^-$ means insertion in lower segment.
\begin{eqnarray}
\mathcal{G}(\tau_1,\tau_2)&=&\frac{1}{N}\sum_i\langle\mathcal{T}_{\tau}(\psi_i(\tau_1)\psi^{\dagger}_i(\tau_2))\rangle\\
G^<(t_1,t_2)&=&-i\,\langle \mathcal{T_C}\left(\psi^+(t_1)\psi^{\dagger}(t_2)\right) \rangle \qquad \text{contour ordered}\nonumber\\
&=&-i\,\langle \mathcal{T_C}\left(\psi(t_1+i\epsilon)\psi^{\dagger}(t_2)\right) \rangle\nonumber\\
&=&i\langle \psi^{\dagger}(t_2)\psi(t_1) \rangle \qquad\qquad \text{operator ordered}\nonumber\\
G^>(t_1,t_2)&=&-i\,\langle \mathcal{T_C}\left(\psi^-(t_1)\psi^{\dagger}(t_2)\right) \rangle \qquad \text{contour ordered}\nonumber\\
&=&-i\,\langle \mathcal{T_C}\left(\psi(t_1-i\epsilon)\psi^{\dagger}(t_2)\right) \rangle\nonumber\\
&=&-i\,\langle \psi(t_1)\psi^{\dagger}(t_2) \rangle \qquad\qquad \text{operator ordered}\nonumber\\
G^R(t_1,t_2)&=&\Theta(t_1-t_2)\left[G^>(t_1,t_2)-G^<(t_1,t_2)\right]\\
\label{GR}
G^A(t_1,t_2)&=&\Theta(t_2-t_1)\left[G^<(t_1,t_2)-G^>(t_1,t_2)\right]\
\label{GA}
\end{eqnarray}
Our convention for the Fourier transforms are
\begin{eqnarray}
f(\omega)=\int_{-\infty}^{\infty} dt \, e^{i\omega t}f(t)\,, && \quad f(t)\,=\int_{-\infty}^{\infty} \frac{d\omega}{2\pi} \, e^{-i\omega t}f(\omega)\nonumber\\
\psi^{\dagger}(\tau)=\sum_{n=-(N-1/2)}^{N-1/2}\frac{e^{i\omega_n\tau}}{\beta}\psi^{\dagger}(\omega_n)\,, && \quad \psi^{\dagger}(\omega_n)=\sum_{n=0}^{N-1}\frac{\beta}{N}\,e^{-i\omega_n\tau}\psi^{\dagger}(\tau)\\
\label{bpsimode}
\psi(\tau)=\sum_{n=-(N-1/2)}^{N-1/2}\frac{e^{-i\omega_n\tau}}{\beta}\psi(\omega_n)\,, && \quad \psi(\omega_n)=\sum_{n=0}^{N-1}\frac{\beta}{N}\,e^{i\omega_n\tau}\psi(\tau)\
\label{psimode}
\end{eqnarray}

\section{Mass versus chemical potential}
\label{massversuschempot}
In many works, mass and chemical potential are confusingly mixed up and used interchangeably. Indeed there are no difference between mass and chemical potential in calculation of thermal partition function using path integral approach in imaginary time). But in real time calculations, mass and chemical potential are very different quantities. While mass term is a part of the Hamiltonian, chemical potential is manifested in the state. So, if one use Wick rotation (analytic continuation) to obtain real time quantities from imaginary time quantities, then the Hamiltonian also includes a mass term. The thermal state also has chemical potential turned on. The starkest difference between chemical potential and mass term can be easily explained in the two free fermionic systems consisting of one fermion each. Consider the two systems to be as follows:
\begin{enumerate}
\item A free Hamiltonian and the system is in a thermal state with chemical potential $\eta$ \footnote{Although the Hamiltonian is zero, this case arises when the bath supplies more fermions or holes depending upon the sign of $\eta$.}.
\begin{equation}
H_1=0, \qquad\qquad Q=\Psi^{\dagger}\Psi, \qquad\qquad Z=\text{Tr}\left( e^{-\beta\eta Q}\right)\
\label{Hfree}
\end{equation}
But the time evolution operator is
\begin{equation}
U=e^{-itH_1}=1
\end{equation}
\item A Hamiltonian consisting of a mass term and the system is a thermal state with only temperature.
\begin{equation}
H_2=\mu \Psi^{\dagger}\Psi\, \qquad\qquad Z=\text{Tr}\left(e^{-\beta H_2}\right)\
\label{Hmass}
\end{equation}
The time evolution operator is
\begin{equation}
U=e^{-itH_2}=e^{-it\mu\Psi^{\dagger}\Psi}\
\end{equation}
\end{enumerate}
The partition functions for the two systems at same temperature $\beta$ are
\begin{eqnarray}
Z_1=\text{Tr}\left(e^{-\beta(H_1+\eta Q}\right) =(1+e^{-\beta\eta})\\
Z_2=\text{Tr}\left(e^{-\beta H_2}\right)=(1+e^{-\beta\mu})\
\end{eqnarray}
Indeed, the partition functions for the two systems are same if one takes $\eta=\mu$ numerically (and at same temperature), but this does not mean dynamically the two systems are same. The Green's function of the first system cannot be obtained by using Wick rotation.
The Green's functions of system 1 in imaginary and real time are given below.
\begin{eqnarray}
\label{gtaumu}
\mathcal{G}_1(\tau_1,\tau_2)&=&\frac{\Theta(\tau_1-\tau_2)e^{-\eta(\tau_1-\tau_2)}-\Theta(-(\tau_1-\tau_2))e^{-\eta(\tau_1-\tau_2+\beta)}}{1+e^{-\beta\eta}}\\
G_1^>(t_1,t_2)&=&-\frac{i}{1+e^{-\beta\eta}}\\
\label{Ggreater1}
G_1^<(t_1,t_2)&=&\frac{i}{1+e^{\beta\eta}}\
\label{Glesser1}
\end{eqnarray}
$\mathcal{G}_1(\tau_1,\tau_2)$ is the Feynman propagator in imaginary time. $G_1^>(t_1,t_2)$ and $G_1^<(t_1,t_2)$ are the greater and lesser Green's functions respectively \footnote{$G^>$ and $G^<$ are not really Green's functions. They are Wightman functions. Their equations of motion are homogeneous.}. Their exact definitions are given in Appendix \ref{conventions}. Using $G^>$ and $G^<$, one can write down the Feynman, retarded and advanced Green's functions. The Green's functions of system 2 are
\begin{eqnarray}
\mathcal{G}_2(\tau_1,\tau_2)&=&\frac{\Theta(\tau_1-\tau_2)e^{-\mu(\tau_1-\tau_2)}-\Theta(-(\tau_1-\tau_2))e^{-\mu(\tau_1-\tau_2+\beta)}}{1+e^{-\beta\mu}}\\
G_2^>(t_1,t_2)&=&-\frac{ie^{-i(t_1-t_2)\mu}}{1+e^{-\beta\mu}}\\
\label{Ggreater2}
G_2^<(t_1,t_2)&=&\frac{ie^{-i(t_1-t_2)\mu}}{1+e^{\beta\mu}}\
\label{Glesser2}
\end{eqnarray}
Note that the real time Green's function for the two systems are related by
\begin{eqnarray}
G_1^{>(<)}(t_1,t_2;\eta=\mu)=e^{it\mu}G_2^{>(<)}(t_1,t_2;\mu)\
\label{Grelation}
\end{eqnarray}
This also holds true for other general Hamiltonians. It can be shown easily using the BCH relations (\ref{BCH1},\ref{BCH2}) and the fact that $\Psi^{\dagger}\Psi$ is a conserved charge so it commutes with the Hamiltonian.


\section{Fluctuation-dissipation theorem with chemical potential}
\label{fdchem}
Consider the lesser Green's function in a thermal ensemble with the chemical potential turned on
\begin{eqnarray}
G^<(-t)&=&i\,\text{Tr}\left[e^{-\beta(H+\eta Q)} \psi^{\dagger}(t)\psi(0))\right]\nonumber\\
&=&i\,\text{Tr}\left[e^{-\beta H-\beta\eta Q} e^{itH}\psi^{\dagger}(0) e^{-itH}\psi(0)\right]\nonumber\\
&=&i\,\text{Tr}\left[e^{i(t+i\beta)H}e^{-\beta\eta Q}\psi^{\dagger}(0)e^{-i(t+i\beta)H}e^{-\beta H}\psi(0)\right]\nonumber\\
&=&i\,\text{Tr}\left[e^{i(t+i\beta)H}\psi^{\dagger}(0) e^{-\beta\eta}e^{-\beta\eta Q}e^{-i(t+i\beta)H}e^{-\beta H}\psi(0)\right]\nonumber\\
&=&-i\,e^{-\beta\eta}\,\text{Tr}\left[\psi^{\dagger}(t+i\beta)e^{-\beta\eta Q}e^{-\beta H}\psi(0)\right]\nonumber\\
&=&i\,e^{-\beta\eta}\,\text{Tr}\left[e^{-\beta H-\beta\eta Q}\psi(0)\psi^{\dagger}(t+i\beta)\right]\nonumber\\
&=&-e^{-\beta\eta}\,G^>(-t-i\beta)
\label{fdt1}
\end{eqnarray}
where in the fourth line we have used the BCH relation (\ref{BCH1}).
Taking $t\to -t$ and Fourier transforming both sides of (\ref{fdt1}) w.r.t. $t$, we have
\begin{eqnarray}
G^<(\omega)=-e^{-\beta\eta-\beta\omega}G^>(\omega)\
\label{dft2}
\end{eqnarray}
The rest of the derivation are the standard steps. One takes the Fourier transform of the retarded Green's function
\begin{eqnarray}
G^R(\omega)&=&\int_{-\infty}^{\infty} dt \, e^{i\omega t} \, \Theta(t)\,\left[G^>(t)-G^<(t)\right]\nonumber\\
&=&\int_{-\infty}^{\infty} \frac{d\omega'}{2\pi} \, \left[1+e^{-\beta(\eta+\omega)}\right]G^>(\omega)\,\frac{i}{\omega-\omega'+i\epsilon}\nonumber\\
\Rightarrow \text{Im}\left[G^R(\omega)\right]&=&-\frac{i}{2} \, \left[1+e^{-\beta(\eta+\omega)}\right]\,G^>(\omega)\nonumber\
\end{eqnarray}
We took $\epsilon\to0^+$ and also note that $G^>(\omega)$ is purely negative imaginary. The real part of $G^R(\omega)$ will be given by the principle value integral. The spectral function is
\begin{eqnarray}
A(\omega)=-2\;\text{Im}\,G^R(\omega)\
\label{defA}
\end{eqnarray}
So, finally we have,
\begin{eqnarray}
\label{fdtfinal1}
G^>(\omega)&=&-\,\frac{i}{1+e^{-\beta(\eta+\omega)}}\,A(\omega)\\
G^<(\omega)&=&\frac{i}{1+e^{\beta(\eta+\omega)}}\,A(\omega)\
\label{fdtfinal2}
\end{eqnarray}
These relations constitute the Fluctuation-Dissipation theorem with chemical potential.

\section{Fermion partition function}
\label{partfunccalc}
In this section we will write down the details involved in the calculation of fermionic partition function and the free energy. We will closely follow \cite{shankarbook}. Consider a one fermion system with the Hamiltonian
\begin{equation}
H_{0}=E\Psi^{\dagger}\Psi
\end{equation}
where $\Psi^{\dagger}$ and $\Psi$ are fermionic operators. Fermionic coherent states are defined as
\begin{eqnarray}
\Psi|\psi\rangle=\psi|\psi\rangle\\
|\psi\rangle=|0\rangle-\psi|1\rangle\\
\langle\bar{\psi}|\Psi^{\dagger}=\langle\bar{\psi}|\bar{\psi}\\
\langle\bar{\psi}|=\langle 0|-\langle 1|\bar{\psi}\
\end{eqnarray}
A complete set of coherent states is given by
\begin{equation}
\mathbf{1}=\int |\psi\rangle \langle \bar{\psi}|\, e^{-\bar{\psi}\psi}d\bar{\psi}d\psi\
\end{equation}
Lastly, for any bosonic operator $\Omega$(consisting of even number of fermionic operators)
\begin{equation}
\text{Tr}\Omega = \int \langle -\bar{\psi}|\Omega|\psi\rangle \,e^{-\bar{\psi}\psi}\,d\bar{\psi}d\psi\
\end{equation}
So we have the partition function
\begin{eqnarray}
Z&=&\text{Tr} \, \,e^{-\beta H}\nonumber\\
&=&\int \langle -\bar{\psi}_0|e^{-\beta H}|\psi_0\rangle e^{-\bar{\psi}_0\psi_0}d\bar{\psi}_0 d\psi_0\
\end{eqnarray}
For finite $\beta$, the expansion of the exponential is not normal ordered. Breaking up $\beta$ into L infinitesimal intervals of size $d\tau$ and inserting complete set of coherent states. Now we get
\begin{eqnarray}
Z&=&\int \langle -\bar{\psi}_0|e^{-d\tau H}|\psi_{L-1}\rangle e^{-\bar{\psi}_{L-1}\psi_{L-1}} \langle -\bar{\psi}_{L-1}|e^{-d\tau H}|\psi_{L-2}\rangle e^{-\bar{\psi}_{L-2}\psi_{L-2}}\times ...\nonumber\\
&&\qquad ...\times\,\langle -\bar{\psi}_2|e^{-d\tau H}|\psi_1\rangle e^{-\bar{\psi_1}\psi_1}\langle -\bar{\psi}_1|e^{-d\tau H}|\psi_0\rangle e^{-\bar{\psi_0}\psi_0} \prod_{i=0}^{L-1}d\bar{\psi}_i d\psi_i\nonumber\\
&=&\int \prod_{i=0}^{L-1} exp \left[\bar{\psi}_{i+1}\psi_i-d\tau E \bar{\psi}_{i+1}\psi_i-\bar{\psi}_i\psi_i\right] d\bar{\psi}_i d\psi_i\
\label{part1}
\end{eqnarray}
where in the second line we have defined $\psi_L=-\psi_0$.
Note that we have used the time and frequency ranges
\begin{eqnarray}
d\tau&=&\frac{\beta}{L},\quad \tau_i \in \left\{0,d\tau,2d\tau,....,(L-1)d\tau\right\},\quad\text{or} \quad \tau_i=i\,d\tau, \, i\in\left\{0,1,2, ... , L-2,L-1\right\}\nonumber\\
d\omega&=&\frac{2\pi}{\beta}, \quad \omega_n=\frac{2\pi}{\beta}\left(n-\frac{1}{2}\right), \, n\in\left\{-\frac{L}{2}+1,\dots,-1,0,1,\dots,\frac{L}{2}\right\}\nonumber\
\end{eqnarray}

In the continuum limit one obtains the well known action of a single fermion.
\begin{eqnarray}
Z&=&\int \prod_{i=0}^{L-1} exp\left[\left\{\frac{\bar{\psi}_{i+1}-\bar{\psi}_i}{d\tau}\psi_i- E \bar{\psi}_{i+1}\psi_i\right\}d\tau\right] d\bar{\psi}_i d\psi_i\nonumber\\
&=&\int \prod_{i=0}^{L-1} exp\left[\int_0^{\beta}d\tau\left\{-\bar{\psi}\partial\psi-E \bar{\psi}\psi\right\}\right] \mathcal{D}\bar{\psi}\mathcal{D}\psi\nonumber\
\end{eqnarray}
where in the second line we have performed an integration by part.

One cannot obtain the correct free energy using the continuum approximation. From the discretized formula, we get
\begin{eqnarray}
Z=Det\begin{bmatrix}
    1 & -1+E\,d\tau & 0 & \dots & 0 \\
    0 & 1 & -1+E\,d\tau & \dots  & 0 \\
    \vdots & \vdots & \vdots & \ddots & \vdots \\
    0 & 0 & \dots & 1  & -1+E\,d\tau \\
    1-E\,d\tau & 0 & 0 & \dots  & 1 \end{bmatrix}
\end{eqnarray}
It can be numerically shown (taking large enough L) that the above expression gives $1+e^{-\beta E}$ which is the well known partition function of a fermion. For a massless free fermion ($E=0$), as expected we get $Z=2$. Note that the non-zero element at the lowermost leftmost corner is important, it fixes the periodic boundary condition, without it the determinant is 1.

%
%

Working in the frequency domain, the continuum formula does not give the correct free energy. For the first/kinetic term, the continuum approximation amounts to replacing $e^{i\omega_n d\tau}-1$ with $i\omega\, d\tau$ while for the second/potential term, the approximation is replacing $e^{i\omega_n d\tau}$ with $1$. But this approximations would fail when $\omega_n$ is very large. So, only quantities which are not sensitive to the high frequencies will be correct. For the exact result, one has to use again the formula (\ref{part1}). Using the mode expansions (\ref{bpsimode},\ref{psimode}), we get
\begin{equation}
Z=\prod_{n=-L/2+1}^{L/2} \left[e^{i\omega_n d\tau}(1-Ed\tau)-1\right]\
\end{equation}
For a massless free fermion ($E=0$),
\begin{equation}
Z=\prod_{n=-L/2+1}^{L/2} \left[e^{i\pi\left(n-1/2\right)/L}-1\right]=2\
\end{equation}
We can see here that the continuum approximation would fail catastrophically in this case.

Working in time domain, the partition function of the complex SYK with mass/charge at inverse temperature $\beta$ is
\begin{eqnarray}
Z&=& Det\begin{bmatrix}
    1+\Sigma(0)d\tau^2 & -1+\mu\,d\tau+\Sigma(d\tau)d\tau^2 & \dots & \Sigma(\beta-d\tau)d\tau^2 \\
    \Sigma(-d\tau)d\tau^2 & 1+\Sigma(0)d\tau^2 & \dots  & \Sigma(\beta-2 d\tau)d\tau^2 \\
    \vdots & \vdots & \ddots & \vdots \\
    \Sigma(-\beta+2 d\tau)d\tau^2 & \dots & 1+\Sigma(0)d\tau^2  & -1+\mu\,d\tau+\Sigma(d\tau)d\tau^2 \\
    1-\mu\,d\tau+\Sigma(-\beta+d\tau)d\tau^2 & \Sigma(-\beta+2 d\tau)d\tau^2 & \dots  & 1+\Sigma(0)d\tau^2 \end{bmatrix}^N\nonumber\\
&& \qquad\qquad  \times\; \text{exp} \left[\frac{N\beta}{4}\sum_{i=-(L-1)}^{(L-1)} d\tau\, \left\{J_2^2 G(i*d\tau) G(-i*d\tau) - \frac{3 J_4^2}{2} \,G(i*d\tau)^2G(-i*d\tau)^2 \right\} \right] \nonumber\\
\label{syknumpart}
\end{eqnarray}
Compare to (\ref{partfunc}), we have performed a coordinate transformation from $(\tau_1,\tau_2)$ to $(\tau_1-\tau_2,\tau_2)$ in the exponential part and we also have used the expression of $\Sigma(\tau_1,\tau_2)$ in (\ref{SD1}).
\bibliography{sykquench} 
\bibliographystyle{JHEP}

\end{document}